\documentclass[11pt,a4paper]{article}
\usepackage[numbers,sort&compress,merge]{natbib}
\usepackage{jheppub}

\usepackage{graphicx}
\usepackage{amsmath,amssymb}
\usepackage{units}
\usepackage{booktabs}

\usepackage{hyperref}

\newcommand{\be}{\begin{equation}}
\newcommand{\ee}{\end{equation}}
\newcommand{\ba}{\begin{eqnarray}}
\newcommand{\ea}{\end{eqnarray}}
\newcommand{\MSb}{$\overline{\text{MS}}$}
\newcommand{\lsim}{\lesssim}
\newcommand{\gsim}{\gtrsim}

\newcommand{\aS}{\alpha_s}
\begin{document}

\title{Higgs boson mass and new physics}

\author[a,b]{Fedor Bezrukov,}
\author[c]{Mikhail~Yu.~Kalmykov,}
\author[c]{Bernd A. Kniehl}
\author[d]{and Mikhail Shaposhnikov}

\affiliation[a]{Physics Department, University of Connecticut,\\
  Storrs, CT 06269-3046, USA}
\affiliation[b]{RIKEN-BNL Research Center, Brookhaven National Laboratory,\\
  Upton, NY 11973, USA}
\affiliation[c]{II. Institut f\"ur Theoretische Physik, Universit\"at Hamburg,\\
  Luruper Chaussee 149, 22761, Hamburg, Germany}
\affiliation[d]{Institut de Th\'eorie des Ph\'enom\`enes Physiques,\\
  \'Ecole Polytechnique F\'ed\'erale de Lausanne, CH-1015 Lausanne,
  Switzerland}

\emailAdd{Fedor.Bezrukov@uconn.edu}
\emailAdd{mikhail.kalmykov@desy.de}
\emailAdd{kniehl@desy.de}
\emailAdd{Mikhail.Shaposhnikov@epfl.ch}

\date{\today}

\abstract{%
  We discuss the lower Higgs boson mass bounds which come from the
  absolute stability of the Standard Model (SM) vacuum and from the
  Higgs inflation, as well as the prediction of the Higgs boson mass
  coming from asymptotic safety of the SM.  We account for the 3-loop
  renormalization group evolution of the couplings of the Standard
  Model and for a part of two-loop corrections that involve the QCD
  coupling $\aS$ to initial conditions for their running.  This is one
  step above the current state of the art procedure (``one-loop
  matching--two-loop running'').  This results in reduction of the
  \emph{theoretical} uncertainties in the Higgs boson mass bounds and
  predictions, associated with the Standard Model physics, to
  $\unit[1\text{--}2]{GeV}$. We find that with the account of existing
  experimental uncertainties in the mass of the top quark and $\aS$
  (taken at $2\sigma$ level) the bound reads $M_H\geq{}M_{\min}$
  (equality corresponds to the asymptotic safety prediction), where
  $M_{\min}=\unit[129\pm6]{GeV}$.  We argue that the discovery of the
  SM Higgs boson in this range would be in agreement with the
  hypothesis of the absence of new energy scales between the Fermi and
  Planck scales, whereas the coincidence of $M_H$ with $M_{\min}$
  would suggest that the electroweak scale is determined by Planck
  physics.  In order to clarify the relation between the Fermi and
  Planck scale a construction of an electron-positron or muon collider
  with a center of mass energy $\sim\unit[200+200]{GeV}$ (Higgs and
  t-quark factory) would be needed.}


\arxivnumber{1205.2893}

\maketitle

%%%%%%%%%%%%%%%%%%%%%%%%%%%%%%%%%%%%%%%%%%%%%%%%%%%%%%%%%%%%%%%%%%%%%%%%
\section{Introduction}
\label{sec:intro}

\begin{figure}
  \centering
  \includegraphics{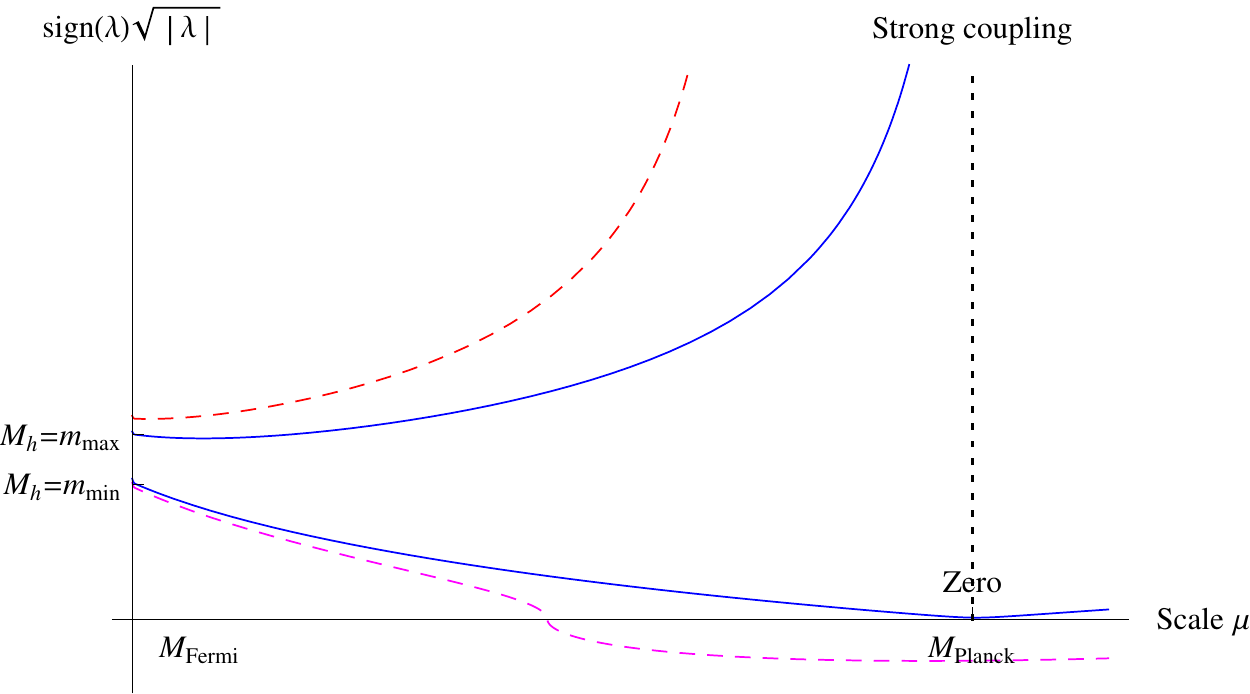}\\[2ex]
  \includegraphics{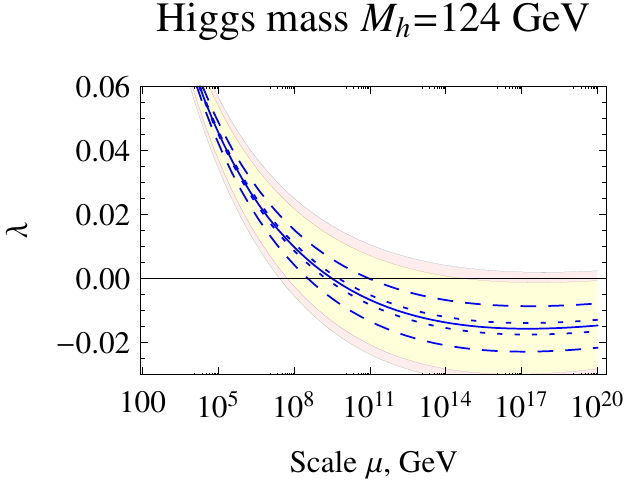} \includegraphics{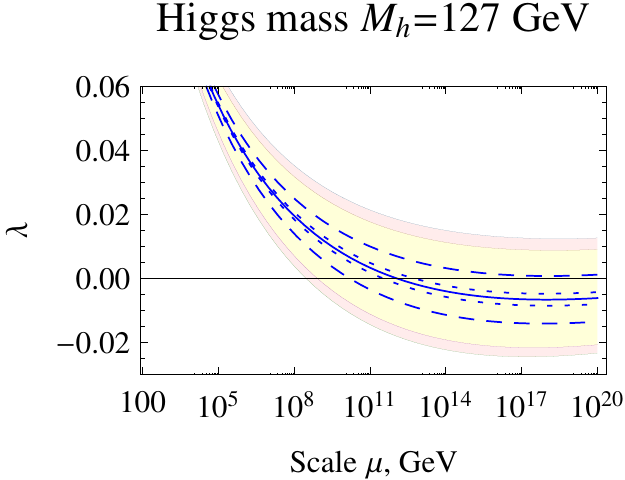}
  \caption{Higgs self-coupling in the SM as a function of the energy
    scale.  The top plot depicts possible behaviors for the whole
    Higgs boson mass range---Landau pole, stable, or unstable
    electroweak vacuum.  The lower plots show detailed behavior for
    low Higgs boson masses, with dashed (dotted) line corresponding to
    the experimental uncertainty in the top mass $M_t$ (strong
    coupling constant $\aS$), and the shaded yellow (pink) regions
    correspond to the total experimental error and theoretical
    uncertainty, with the latter estimated as $\unit[1.2]{GeV}$
    ($\unit[2.5]{GeV}$), see section \ref{sec:Stability} for detailed
    discussion.}
  \label{fig:lambda}
\end{figure}

The mass $M_H$ of the Higgs boson in the Standard Model is an
important indicator of the presence of new energy scales in particle
physics.  It is well known that if $M_{\min}^{\text{meta}} < M_H
< M_{\max}^{\text{Landau}}$ then the SM is a consistent effective
field theory all the way from the Fermi scale up to the (reduced)
Planck scale $M_P=\unit[2.44\times10^{18}]{GeV}$.  The upper limit
comes from the requirement that the Landau pole in the scalar
self-coupling\footnote{To be more precise, the scalar self-coupling is
infinite in the one-loop approximation only.  If higher order terms
are included, it may not become infinite, but move away from the
region of the weak coupling.} must not appear at energies below $M_P$
\cite{Maiani:1977cg,Cabibbo:1979ay,Lindner:1985uk}.  The lower limit
comes from the requirement of the stability of the SM vacuum against
tunneling to the states with the Higgs field $\phi$ exceeding
substantially the electroweak value
$\unit[250]{GeV}$~\cite{Krasnikov:1978pu,Hung:1979dn,Politzer:1978ic}
(see Fig.~\ref{fig:lambda}).

The estimates of $M_{\max}^{\text{Landau}}$ give a number around
$\unit[175]{GeV}$
\cite{Maiani:1977cg,Cabibbo:1979ay,Lindner:1985uk,Hambye:1996wb} which
is in the $M_H$ range excluded (at least in the range
$\unit[129-525]{GeV}$) by the searches for the SM Higgs boson at the
LHC and Tevatron \cite{Chatrchyan:2012tx,ATLAS:2012ae}.  In other
words, we already know that the SM is a weakly coupled theory up to
the Planck scale.  

One can distinguish between two types of the stability bounds. If
$M_H>M_{\min}^{\text{stability}}$, the electroweak vacuum is
absolutely stable, whereas if $M_{\min}^{\text{meta}} < M_H <
M_{\min}^{\text{stability}}$, then it is metastable with the
life-time exceeding that of the Universe. Numerically,
$M_{\min}^{\text{meta}}\simeq \unit[111]{GeV}$ \cite{Espinosa:2007qp}. The
existence of the Higgs boson with the mass smaller than
$M_{\min}^{\text{meta}}$ would provide an undisputed argument in favor
of existence of new physics between the Fermi and Planck scale.
However, already since LEP we know that this is not the case.

The Higgs mass $M_{\min}^{\text{stability}}$ is not at all special
from the point of view of the validity of the SM up to the Planck
scale. The value of  $M_{\min}^{\text{stability}}$, however, plays a
crucial role if the Standard Model is embedded to a bigger picture
which includes gravity. First, only if  $M_H >
M_{\min}^{\text{stability}}$, the Higgs boson of the SM can play the
role of the inflaton and make the Universe flat,  homogeneous and
isotropic, and produce a necessary spectrum of perturbations needed
for structure formation \cite{Bezrukov:2007ep,Bezrukov:2009db}.
Second,  $M_H = M_{\min}^{\text{stability}}$ is a prediction of
asymptotically safe scenario for the SM \cite{Shaposhnikov:2009pv},
making it consistent up to arbitrary large scale.

Thus, we will focus on the upgrade of existing computations of
$M_{\min}^{\text{stability}}$ and on the discussion of the
significance of the relation between the Higgs boson (to be discovered
yet) mass $M_H$ and $M_{\min}^{\text{stability}}$ for beyond the SM
(BSM) physics.

The computation of $M_{\min}^{\text{stability}}$ has been already done
in a large number of papers
\cite{Altarelli:1994rb,Casas:1994qy,Casas:1996aq,Espinosa:2007qp,
Bezrukov:2009db,Holthausen:2011aa}.  It is divided into two parts. 
The first one is the determination of the \MSb\ parameters from the
physical observables and the second one is the renormalization group
running of the \MSb\ constants from the electroweak to a high energy
scale.  The most advanced recent works
\cite{Espinosa:2007qp,Bezrukov:2009db,Holthausen:2011aa,EliasMiro:2011aa}
use the so-called ``one-loop-matching--two-loop-running'' procedure. 
It can determine the Higgs boson mass bounds with the theoretical
accuracy of $\unit[2-5]{GeV}$ (see the discussion of uncertainties in
\cite{Bezrukov:2009db} and below).  Meanwhile, the most important
terms in the 3-loop running of the gauge and Higgs coupling constants
were computed in \cite{Mihaila:2012fm,Chetyrkin:2012rz} (we thank
K.~Chetyrkin and M.~Zoller for sharing these results with us prior to
publication). The present work accounts for $O(\alpha\aS)$ corrections
in the \MSb-pole matching procedure, which were not known previously.
This allows us to decrease the theoretical uncertainties in the Higgs
boson mass prediction/bounds, associated with the SM physics down to
$\unit[1\text{--}2]{GeV}$. This is a new result, based on a superior
partial ``two-loop-matching--three-loop-running'' procedure.  These
findings are described in Section \ref{sec:Results}.  We will see that
the experimental errors in the mass of the top-quark and in the value
of the strong coupling constant are too large to settle up the
question of the stability of the electroweak vacuum, even if the LHC
will confirm the evidence for the Higgs signal presented by the ATLAS
and CMS collaborations \cite{ATLAS:2012ae,Chatrchyan:2012tx} in the
region $M_H=\unit[124-126]{GeV}$.

In Section \ref{sec:BSM} we will discuss the significance of the
relationship between the true Higgs boson mass $M_H$ and
$M_{\min}^{\text{stability}}$ for BSM physics.  We will argue that if
$M_H=M_{\min}^{\text{stability}}$ then the electroweak symmetry
breaking is likely to be determined by Planck physics and that this
would indicate an absence of new energy scales between the Fermi and
gravitational scales.  We will also address here the significance of
$M_{\min}^{\text{stability}}$ for the SM with gravity included.  Of
course, this can only be done under certain assumptions. 
Specifically, we will discuss the non-minimal coupling of the Higgs
field to the Ricci scalar (relevant for
Higgs-inflation~\cite{Bezrukov:2007ep,Bezrukov:2008ej,Bezrukov:2009db})
and the asymptotic safety scenario for the
SM~\cite{Shaposhnikov:2009pv}.

In Section \ref{sec:concl} we present our conclusions.  We will argue
that if only the Higgs boson with the mass around
$M_{\min}^{\text{stability}}$ and nothing else will be found at the
LHC, the next step in high energy physics should be the construction a
new electron-positron (or muon) collider---the Higgs and t-factory. 
It will not only be able to investigate in detail the Higgs and top
physics, but also elucidate the possible connection of the Fermi and
Planck scales.

Appendix~\ref{app:PoleMatching} contains a full set of formulas
required for the determination of the \MSb\ coupling constants from
the pole masses of the SM particles, including the corrections of the
orders of up to $O(\aS^3)$, $O(\alpha)$, and $O(\alpha\aS)$.  The
computer code for the matching is made publicly available at
\url{http://www.inr.ac.ru/~fedor/SM/}.

%%%%%%%%%%%%%%%%%%%%%%%%%%%%%%%%%%%%%%%%%%%%%%%%%%%%%%%%%%%%%%%%
%%%%%%%%%%%%%%%%%%%%%%%%%%%%%%%%%%%%%%%%%%%%%%%%%%%%%%%%%%%%%%%%

\section{The stability bound}
\label{sec:Stability}

\begin{figure}
  \includegraphics[width=\textwidth]{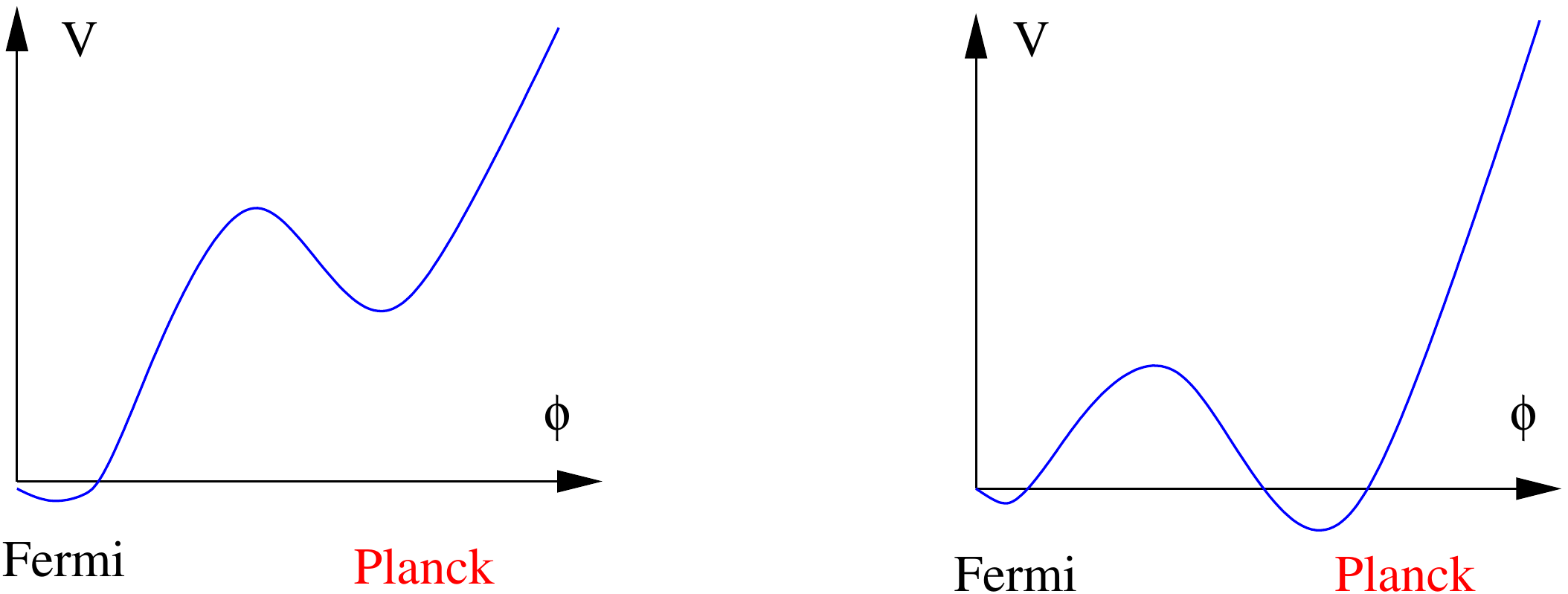}
  \caption{Schematic depiction of the SM effective potential $V$ for
    the Higgs field for $M_H>M_{\min}^{\text{stability}}$ (left) and
    $M_H<M_{\min}^{\text{stability}}$ (right).}
  \label{stability}
\end{figure}

The stability bound will be found in the ``canonical'' SM, without any
new degrees of freedom or any extra higher dimensional operators
added, see Fig.~\ref{stability}.

%%%%%%%%%%%%%%%%%%%%%%%%%%%%%%%%%%%%%%%%%%%%%%%%%%%%%%%%%%%%%%%%%%%%%
%%%%%%%%%%%%%%%%%%%%%%%%%%%%%%%%%%%%%%%%%%%%%%%%%

\subsection{The benchmark mass}
\label{sec:mmin}

It will be convenient for computations to introduce yet another
parameter, ``benchmark mass'', which we will call $M_{\min}$ (without
any superscript). Suppose that all parameters of the SM, except for
the Higgs boson mass, are exactly known.  Then $M_{\min}$, together
with the normalisation point $\mu_0$, are found from the solution of
two equations:
\be
  \lambda(\mu_0)=0,  \quad  \beta_\lambda(\lambda(\mu_0))=0,
  \label{def}
\ee
where $\beta_\lambda$ is the $\beta$-function governing the
renormalisation group (RG) running of $\lambda$.  Here we define all
the couplings of the SM in the \MSb\ renormalisation scheme which is
used de-facto in the most of the higher-loop computations. Clearly, if
any other renormalization scheme is used, the equations
$\lambda=\beta_\lambda=0$ will give another benchmark mass, since the
definition of all the couplings are scheme dependent.

The procedure of computing $M_{\min}$ is very clean and transparent. 
Take the standard \MSb\ definition of all coupling constants of the
SM, fix all of them at the Fermi scale given the experimentally known
parameters such as the mass of the top quark, QCD coupling, etc., and
consider the running Higgs self-coupling $\lambda(\mu)$ depending on
the standard t'Hooft-Veltman parameter $\mu$.  Then, adjust $M_{\min}$
in such a way that equations (\ref{def}) are satisfied at some
$\mu_0$.

For the stability bound one should find the effective potential
$V(\phi)$ and solve the equations
\be
  V(\phi_{SM})=V(\phi_1),  \quad  V'(\phi_{SM})=V'(\phi_1)=0,
  \label{pot}
\ee
where $\phi_{SM}$ corresponds to the SM Higgs vacuum, and $\phi_1$
correspond to the extra vacuum states at large values of the scalar
field.  Though the effective potential and the field $\phi$ are both
gauge and scheme dependent, the solution for the Higgs boson mass to
these equations is gauge and scheme invariant.

In fact, $M_{\min}^{\text{stability}}$ is very close to $M_{\min}$.
Numerically, the difference between them is much smaller, than the
current theoretical and experimental precisions for $M_{\min}$, see
below.  The following well known argument explains why this is the
case. The RG improved effective potential for large $\phi$ can be
written as~\cite{Coleman:1973jx,Casas:1994qy,Casas:1996aq}
\be
  V(\phi) \propto \lambda(\phi)\phi^4 \left[
    1+O\left( \frac{a}{4\pi}\log(M_i/M_j) \right)
  \right],
\ee
where $a$ is here the common name for all the SM coupling constants
(which are rather small at Planck scale), and $M_i$ are the masses of
different particles in the background of the Higgs field.  If
$O(\alpha)$ corrections are neglected, the equations (\ref{pot})
coincide with (\ref{def}), meaning that $M_{\min}\simeq
M_{\min}^{\text{stability}}$.  The numerical evaluation for one loop
effective potential gives $\Delta{}m^{\text{stability}} \equiv
M_{\min}^{\text{stability}}-M_{\min} \simeq \unit[-0.15]{GeV}$, which
can be neglected in view of uncertainties discussed below.

Note that in many papers the stability bound is shown as a function of
the cutoff scale $\Lambda$ (the energy scale up to which the SM can be
considered as a valid effective field theory).  It is required that
$V(\phi)>V(\phi_{SM})$ for all $\phi<\Lambda$.  This can be
reformulated as $\lambda(\mu)>0$ for all $\mu<\Lambda$ with pretty
good accuracy. Interestingly, if $\Lambda=M_P$, this bound is very
close to the stability bound following from eq.~(\ref{pot}), having
nothing to do with the Planck scale (see also below).  Note also that
the uncertainties in experimental determinations of $M_t$ and $\aS$
together with theoretical uncertainties, described in the next
section, lead to significant changes in the scale $\Lambda$. 
Fig.~\ref{fig:lambda} illustrates that for Higgs boson masses
$124-127$ GeV this scale may vary from $\unit[10^8]{GeV}$ up to
infinity within currently available precisions.

\subsection{Value of $M_{\min}$}
\label{sec:Results}

The state of art computation of $M_{\min}$ contained up to now the so
called one-loop \MSb-pole matching, relating the experimentally
measured physical parameters to the parameters of the SM in the \MSb\
subtraction scheme (to be more precise, the two-loop $\aS$ corrections
to the top pole mass--\MSb\ mass relation has been included
\cite{Espinosa:2007qp}).  Then the results of the first step are
plugged into two-loop RG equations and solved numerically.

Before discussing the upgrade of the
one-loop-matching--two-loop-running procedure, we will remind of the
results already known and their uncertainties.  We will make use of
our computations of $M_{\min}$ presented in
\cite{Bezrukov:2009db}.\footnote{The main interest in this paper was
the lower bound on the Higgs boson mass in the Higgs-inflation, see
below.  However, $M_{\min}$ has been estimated as well as a by-product
of the computation.}  A somewhat later papers
\cite{Ellis:2009tp,Holthausen:2011aa,EliasMiro:2011aa} contains
exactly the same numbers for $M_{\min}^{\text{stability}}$ (note,
however, that the theoretical uncertainties were not discussed in
\cite{Ellis:2009tp}).  See also earlier computations in
\cite{Altarelli:1994rb,Casas:1994qy,Casas:1996aq,Hambye:1996wb,Espinosa:2007qp,
EliasMiro:2011aa}.

In \cite{Bezrukov:2009db} we found:
\be
  M_{\min}= \left[ 126.3
    + \frac{M_t-\unit[171.2]{GeV}}{\unit[2.1]{GeV}} \times 4.1
    - \frac{\aS-0.1176}{0.002} \times 1.5
  \right]\unit[{}]{GeV}, 
  \label{pred}
\ee 
and estimated the theoretical uncertainties as summarized in
Table~\ref{tb:olderrors} (see also
\cite{Holthausen:2011aa}).\footnote{The reader not interested in
details of the comparison with the previous results can skip this
discussion and restart at the following paragraph.}  While repeating
this analysis we found some numerical errors which are given at the
bottom section of this table (see a detailed discussion below).  In
total, they shift the value given in eq.~(\ref{pred}) up by
$\unit[0.7]{GeV}$.  As for uncertainties, they were estimated as
follows.  The one-loop matching formulas can be used directly at
$\mu=m_t$, or at some other energy scale, e.g.\ at $\mu=M_Z$, and then
the coupling constants at $m_t$ can be derived with the use of RG
running.  The difference in procedures gives an estimate of two-loop
effects in the matching procedure. This is presented by the first two
lines in Table \ref{tb:olderrors} (in fact, we underestimated before
the uncertainty from $\lambda$ matching---previously we had here
$\unit[1.2]{GeV}$ and now $\unit[1.7]{GeV}$).  The next two lines are
associated with 3 and 4-loop corrections to the top Yukawa $y_t$.  The
3-loop corrections were computed in
\cite{Chetyrkin:1999ys,Chetyrkin:1999qi,Melnikov:2000qh} and the
four-loop $\aS$ contribution to the top mass was guessed to be of the
order $\delta y_t(m_t)/y_t(m_t) \simeq 0.001$ in
\cite{Kataev:2009ns,Kataev:2010zh}.  The non-perturbative QCD effects
in the top pole mass--\MSb\ mass matching are expected to be at the
same level \cite{Smith:1996xz,Hoang:2008yj,Hoang:2009yr}.  For 3-loop
running we put the typical coefficients in front of the largest
couplings $\aS$ and $y_t$.  If these uncertainties are not correlated
and can be summed up in squares, the theoretical uncertainty is
$\unit[2.5]{GeV}$.  If they are summed up linearly, then the
theoretical error can be as large as $\sim\unit[5]{GeV}$.

\begin{table}
  \begin{center}
    \begin{tabular}{lcc}
  \toprule %============================================================================
  Source of uncertainty        & Nature of estimate                  &
                                                   $\Delta_{\text{theor}}M_{\min}$, GeV \\
  \midrule %----------------------------------------------------------------------------
  2-loop matching $\lambda$    & Sensitivity to $\mu$                & 1.7 \\
  2-loop matching $y_t$        & Sensitivity to $\mu$                & 0.6 \\
  3-loop $\aS$ to $y_t$        & known                               & 1.4 \\
  4-loop $\aS$ to $y_t$        & educated guess \cite{Kataev:2009ns,Kataev:2010zh} & 0.4 \\
  confinement, $y_t$           & educated guess $\sim\Lambda_{QCD}$  & 0.5 \\
  3-loop running $M_W \to M_P$ & educated guess                      & 0.8 \\
  \midrule %----------------------------------------------------------------------------
  total uncertainty           & sum of squares                       & 2.5 \\
  total uncertainty           & linear sum                           & 5.4 \\
  \bottomrule %=========================================================================
  \toprule %============================================================================
  Corrections to~\cite{Bezrukov:2009db}
                              &                                      &
                                                                 $\Delta M_{\min}$, GeV \\
  \midrule %----------------------------------------------------------------------------
  Typos in the code used in~\cite{Bezrukov:2009db}
                              & error                                & +0.2 \\
  Extra $\delta_{\mathrm{t}}^{\text{QED}}$ in (A.5) of~\cite{Espinosa:2007qp}
                              & error                                & +0.4 \\
  ``Exact'' formula instead of \\
  approximation (2.20) in~\cite{Hempfling:1994ar}
                              & clarification                        & +0.1 \\
  \midrule %----------------------------------------------------------------------------
  Total correction to (7.1) of~\cite{Bezrukov:2009db}
                              &                                      & +0.7 \\
  \midrule %----------------------------------------------------------------------------
  \multicolumn{2}{l}{Total shift to be applied to (7.1)
    of~\cite{Bezrukov:2009db} for comparison}                        &   +0.7 \\
  \bottomrule %=========================================================================
    \end{tabular}
  \end{center}
  \caption{Theoretical uncertainties and mistakes in the $M_{\min}$ evaluation
    in~\cite{Bezrukov:2009db}.}
  \label{tb:olderrors}
\end{table}

Now, this computation can be considerably improved.  First, in
\cite{Mihaila:2012fm} the 3-loop corrections to the running of all
gauge couplings has been calculated.  Second,
in~\cite{Chetyrkin:2012rz} the leading contributions (containing the
top Yukawa and $\aS$) to the running of the top quark Yukawa and the
Higgs boson self coupling have been determined.  This removes the
uncertainty related to 3-loop RG running.  In addition, in the present
paper, we determine the two-loop corrections of the order of
$O(\alpha\aS)$ to the matching of the pole masses and the top quark
Yukawa and Higgs boson self coupling constants.  Also, the
known~\cite{Chetyrkin:1999ys,Chetyrkin:1999qi,Melnikov:2000qh} three
loop QCD correction to the top quark mass relation of the order
$O(\aS^3)$ can be included (previously it had been used for estimates
of uncertainties).  All this considerably decreases the theoretical
uncertainties in $M_{\min}$.

The individual contributions of the various new corrections on top of
the previous result are summarized in the Table~\ref{tab:dm}.  It is
clearly seen that there are two new significant contributions---one is
the three-loop pure QCD correction to the top quark
mass~\cite{Chetyrkin:1999ys,Chetyrkin:1999qi,Melnikov:2000qh}, and
another is the two loop correction $O(\alpha\aS)$ to the Higgs boson
mass, found in the present paper. Together the new contributions sum
to the overall shift of the previous prediction \cite{Bezrukov:2009db}
by $\unit[-0.89]{GeV}$, giving the result\footnote{Note, that this
value is the benchmark mass defined by eq.~(\ref{def}).  The mass,
corresponding to the metastability bound (\ref{pot}) is
$\sim\unit[0.1\text{--}0.2]{GeV}$ smaller.}
\begin{equation}
  \label{eq:mmin}
  M_{\min} = \left[
    128.95
    + \frac{M_t-\unit[172.9]{GeV}}{\unit[1.1]{GeV}} \times 2.2
    - \frac{\aS-0.1184}{0.0007} \times 0.56
  \right] \unit{GeV}.
\end{equation}

\begin{table}
  \begin{center}
    \begin{tabular}{lc}
      \toprule %================================================================
      Contribution                         & $\Delta M_{\min}$, GeV \\
      \midrule %----------------------------------------------------------------
      Three loop beta functions            & -0.23 \\
      $\delta y_t\propto O(\aS^3)$         & -1.15 \\
      $\delta y_t\propto O(\alpha\aS)$     & -0.13 \\
      $\delta \lambda\propto O(\alpha\aS)$ & 0.62 \\
      \bottomrule %=============================================================
    \end{tabular}
  \end{center}
  \caption{Contributions to the value of the $M_{\min}$.}
  \label{tab:dm}
\end{table}

The new result (\ref{eq:mmin}) is less than $\unit[0.2]{GeV}$ away
from the old one (\ref{pred}) if the same central values for $M_t$ and
$\aS$ are inserted.  This coincidence is the result of some magic.  In
the old evaluation several mistakes were present, summarized in
Table~\ref{tb:olderrors}.  The largest one was the double counting of
$\delta_t^{\text{QED}}$ in (A.5) of~\cite{Espinosa:2007qp}, as
compared to the original result~\cite{Hempfling:1994ar}.  Also, there
were minor typos in the computer code for the matching of the Higgs
coupling constant, and finally there was a small correction coming
from the use of an approximate rather than exact one loop formula for
$O(\alpha)$ corrections from~\cite{Hempfling:1994ar}.  These
corrections add $\unit[0.7]{GeV}$ to the original number in
\cite{Bezrukov:2009db}.  By chance this almost exactly canceled the
$\unit[-0.89]{GeV}$ contribution from the higher loops,
Table~\ref{tab:dm},  nearly leading to a coincidence of
(\ref{eq:mmin}) and (\ref{pred}).

Table \ref{tb:newerrors} summarizes the uncertainties in the new
computation.  It contains fewer lines.  Now we can ignore safely the
error from higher order (4-loop) RG corrections for the running up to
the Planck scale.  The first two lines were derived in the same manner
as previously.  For the Higgs boson self-coupling we can use the
matching formulas (\ref{higgs:12L}) to get the value of $\lambda(\mu)$
at scale $\mu=M_t$ directly, or to get the value $\lambda(M_Z)$ and
then evolve the constants to the scale $\mu=M_t$ with the RG
equations.  The obtained difference $\delta\lambda(M_t)/\lambda(M_t)
\simeq 0.016 $ corresponds to the error $\delta
m\sim\unit[1.0]{GeV}$.  A similar procedure of comparing evolution
between $M_t$ and $M_Z$ using RG equations and direct matching
formulas to the order $O(\aS^3,\alpha,\alpha\aS)$ leads for the chang
in the top quark Yukawa $\delta y_t/y_t\sim0.0005$, leading to $\delta
M\sim\unit[0.2]{GeV}$.  Note, however, that strictly speaking this
test verifies the error of the $\mu$ dependent terms in the matching
formulas, while the constant ones may lead to larger contributions. 
We also do not estimate now the contributions of the order
$O(\alpha^2)$, where formal order in $\alpha$ may correspond to
$y_t^4$.  Thus, this estimate should be better considered as a lower
estimate of the error.  The 4-loop matching and confinement
contributions are the same as before.

\begin{figure}
  \centering
  \includegraphics{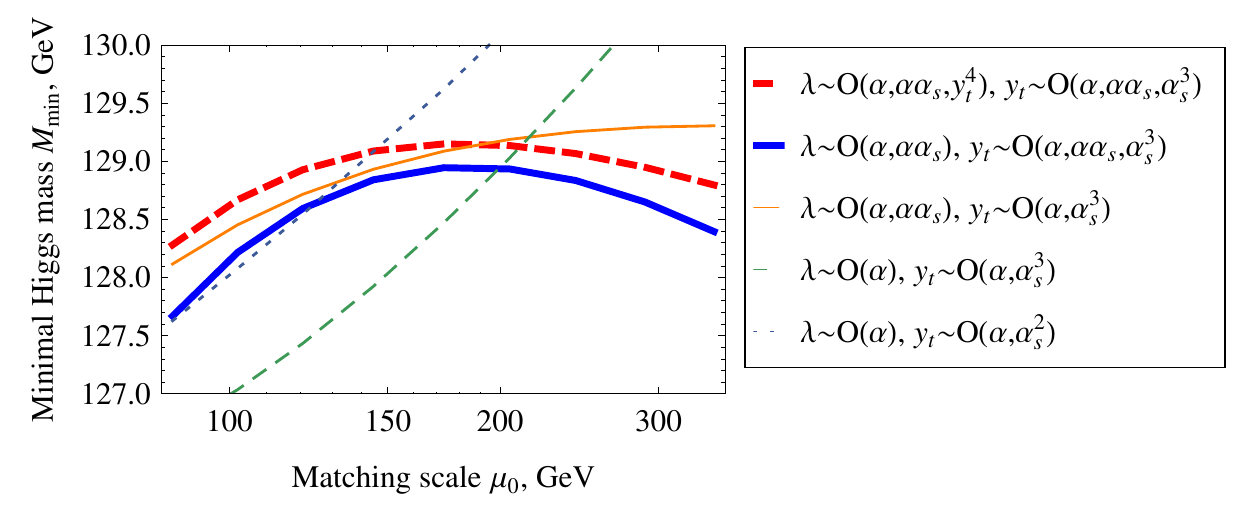}
  \caption{The dependence of the reference Higgs boson mass $M_{\min}$
    on the matching scale $\mu_0$ (the \MSb\ constants are obtained by
    matching formulas at scale $\mu_0$ and then used for the solution
    of the equations (\ref{def})).  The thick solid line corresponds
    to the full matching formulas $\lambda\sim O(\alpha,\alpha\aS)$,
    $y_t\sim O(\aS^3,\alpha,\alpha\aS)$; the thin lines correspond to
    using matching formulas of lower order.  The thick dashed line
    corresponds to using additionally the two loop electroweak
    contributions to the higgs coupling constant in the gauge-less
    limit, eq.~(48) of \cite{Degrassi:2012ry}, see discussion in
    ``Note added''.  Here
    $M_t=\unit[172.9]{GeV}$ and $\aS=0.1184$.}
  \label{fig:mu0dep}
\end{figure}

As an indication of the dependence on the matching point we present
Fig.~\ref{fig:mu0dep}, where the reference Higgs boson mass $M_{\min}$
was obtained using the matching formulas at scale $\mu_0$ varying
between the Z-boson and top quark masses.  One can see that the
overall change of the Higgs boson mass is about GeV.

\begin{table}
  \begin{center}
    \begin{tabular}{lcc}
  \toprule %============================================================================
  Source of uncertainty        & Nature of estimate                  &
                                                   $\Delta_{\text{theor}}M_{\min}$, GeV \\
  \midrule %----------------------------------------------------------------------------
  3-loop matching $\lambda$    & Sensitivity to $\mu$                & 1.0 \\
  3-loop matching $y_t$        & Sensitivity to $\mu$                & 0.2 \\
  4-loop $\aS$ to $y_t$        & educated guess \cite{Kataev:2009ns,Kataev:2010zh} & 0.4 \\
  confinement, $y_t$           & educated guess $\sim\Lambda_{QCD}$  & 0.5 \\
  4-loop running $M_W \to M_P$ & educated guess                      & $<0.2$ \\
  \midrule %----------------------------------------------------------------------------
  total uncertainty            & sum of squares                      & 1.2 \\
  total uncertainty            & linear sum                          & 2.3 \\
  \bottomrule %=========================================================================
    \end{tabular}
  \end{center}
  \caption{Theoretical uncertainties in the present $M_{\min}$ evaluation.}
  \label{tb:newerrors}
\end{table}

If we assume that these uncertainties are not correlated and symmetric
we get a theoretical error in the determination of the critical Higgs
boson mass $\delta{}m_{\text{theor}} \simeq \unit[1.2]{GeV}$. If they
are summed up linearly, we get an error of $\unit[2.4]{GeV}$. We leave
it to the reader to decide which estimate of the uncertainties is more
adequate. The precision of the theoretical value of $M_{\min}$ can be
further increased by computing the $O(\alpha^2)$ two-loop corrections
to the matching procedure. Numerically, the most important terms are
those when $\alpha$ corresponds to $y_t^2$ and $\lambda$.

\begin{figure}
  \centering
  \includegraphics{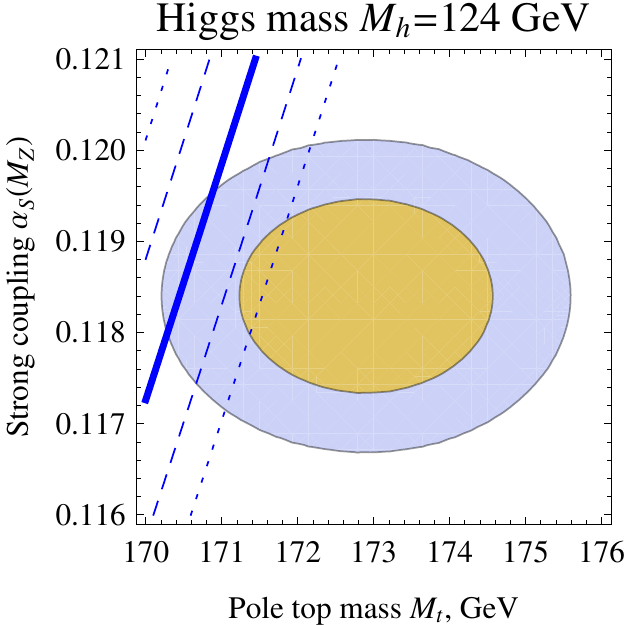}
  \includegraphics{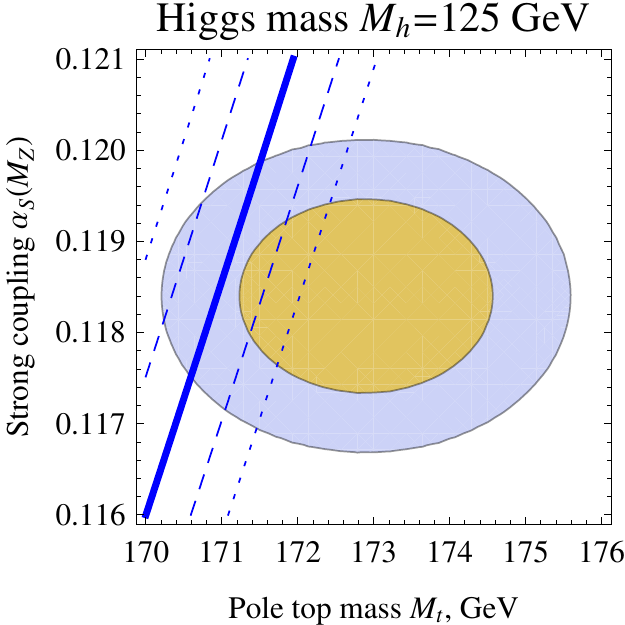}\\[2ex]
  \includegraphics{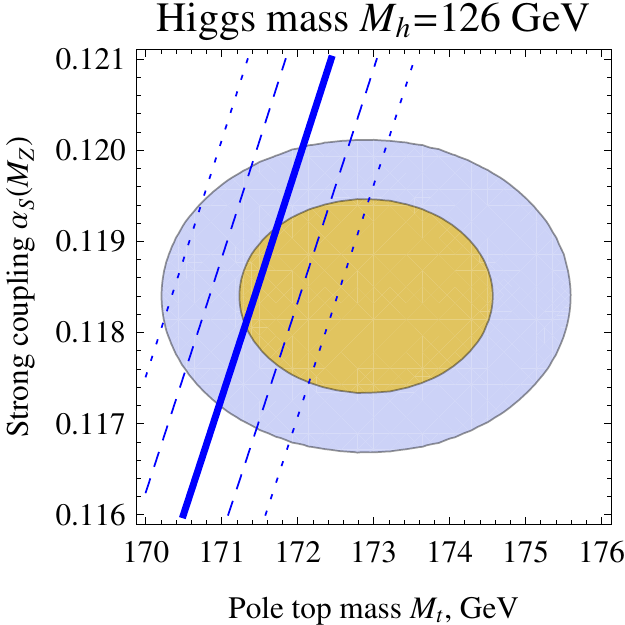}
  \includegraphics{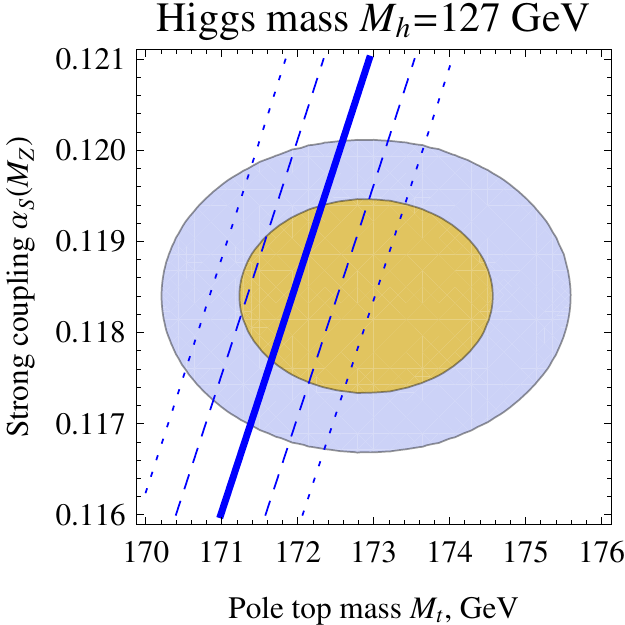}
  \caption{The values of the strong coupling constant $\aS$ and top mass $M_t$
    corresponding to several minimal Higgs boson mass $M_{\min}$.  The 68\% and
    95\% experimentally allowed regions for $\aS$ and $M_t$ are given by shaded
    areas.  The dashed (dotted) lines correspond to $\unit[1.2]{GeV}$
    ($\unit[2.45]{GeV}$) uncertainty in the $M_{\min}$ theoretical
    determination.}
  \label{fig:asmt}
\end{figure}

The result (\ref{eq:mmin}) is visualized by Fig.~\ref{fig:asmt}.  The
experimentally allowed regions for the top mass $M_t$ and strong
coupling $\aS$ are adopted PDG 2010 edition
\cite{PDG2010}.\footnote{Note however, that the current experimental
error estimate is based on averaging over different experimental
approaches.  In some methods quite a different central values are
obtained.  See e.g.\ \cite{Alekhin:2012ig,Watt:2011kp,Watt:2012fj}
about $\aS$ determination
and~\cite{Langenfeld:2009wd,Lancaster:2011wr,Abazov:2011pta} about
$M_t$.}  On top of these allowed regions the bands corresponding to
the reference values of the Higgs boson mass $M_{\min}$ being equal to
$\unit[124,125,126,127]{GeV}$ are shown, with the dashed and dotted
lines corresponding to quadratically or linearly added estimates of
theoretical uncertainties.

One can see that the accuracy of theoretical computations and of the
experimental measurements of the top and the Higgs boson masses does
not allow yet to conclude with confidence whether the discovery of the
Higgs boson with the mass $\unit[124-127]{GeV}$ would indicate
stability or metastability of the SM vacuum.  All these reference
values of Higgs masses are compatible within $2\sigma$ with current
observations.

\section{$M_{\min}$ and BSM physics}
\label{sec:BSM}

\begin{figure}
  \centering
  \includegraphics{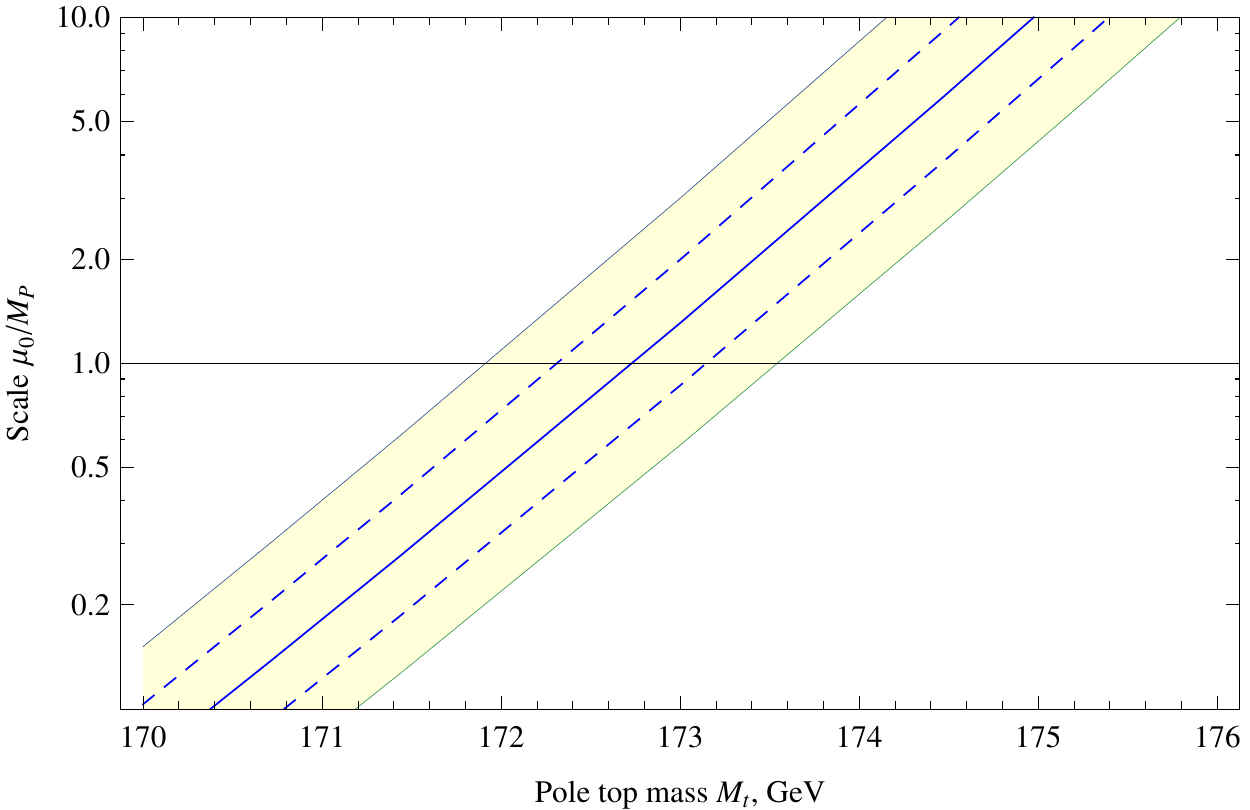}
  \caption{The scale $\mu_0$ (solution of (\ref{def})) depending on
    the top mass $M_t$.  The dashed lines correspond to $1\sigma$
    uncertainty in the $\aS$. The yellow shaded region corresponds to
    adding the $\aS$ experimental error and the theoretical
    uncertainty in the matching of the top Yukawa $y_t$ and top pole
    mass.}
  \label{fig:mu0}
\end{figure}

Our definition of the ``benchmark'' Higgs boson mass consists of the
solution of the two equations~(\ref{def}) and gives, in addition to
$M_{\min}$, the value of the scale $\mu_0$ at which the scalar
self-coupling and its $\beta$-function vanish simultaneously.  The
central value for $\mu_0$ is $\unit[2.9\times10^{18}]{GeV}$ and is
quite stable if $m_t$ and $\aS$ are varied in their confidence
intervals (see Fig.~\ref{fig:mu0}).  One can see that there is a
remarkable coincidence between $\mu_0$ and the (reduced) Planck scale
$M_P=\unit[2.44\times10^{18}]{GeV}$.  The physics input in the
computation of $\mu_0$ includes the parameters of the SM only, while
the result gives the gravity scale.  A possible explanation may be
related to the asymptotic safety of the SM, see
\cite{Shaposhnikov:2009pv} and below.\footnote{Of course, this is not
the only way to predict the Higgs boson mass close to $M_{\min}$. Cf.\
the ``multiple point principle''
of~\cite{Bennett:1993pj,Froggatt:1995rt}, requiring the degeneracy
between the SM vacuum and an extra one appearing at the Planck scale,
the inflation from false vacuum
decay~\cite{Isidori:2007vm,Masina:2011aa,Masina:2011un,Masina:2012yd},
etc.}  It remains to be seen if this is just the random play of the
numbers or a profound indication that the electroweak symmetry
breaking is related to Planck physics.  If real, this coincidence
indicates that there should be no new energy scales between the Planck
and Fermi scales, as they would remove this coincidence unless some
conspiracy is taking place.

We will discuss below two possible minimal embeddings of the SM to the
theory of gravity and discuss the significance of $M_{\min}$ in them.

\subsection{Asymptotic safety}
\label{sec:AS}

The asymptotic safety of the SM \cite{Shaposhnikov:2009pv}, associated
with the asymptotic safety of gravity \cite{Weinberg1979}, is strongly
related to the value of the Higgs boson mass.  Though General
Relativity is non-renormalizable by perturbative methods, it may exist
as a field theory non-perturbatively, exhibiting a non-trivial
ultraviolet fixed point (for a review see \cite{Niedermaier:2006wt}). 
If true, all other coupling of the SM (including the Higgs
self-interaction) should exhibit an asymptotically safe behaviors with
the gravity contribution to the renormalisation group running
included.

\begin{figure}
  \includegraphics[width=\textwidth]{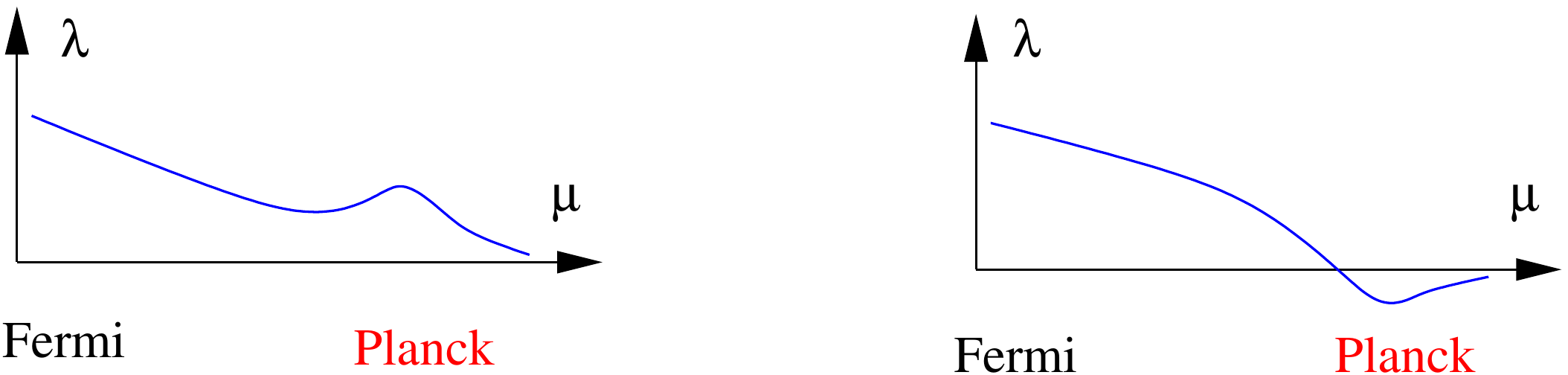}
  \caption{Schematic depiction of the behavior of the scalar
    self-coupling if $a_h<0$ for $M_{\min}^{\text{safety}} < M_H <
    M_{\max}^{\text{safety}}$ (left) and $M_H<
    M_{\min}^{\text{safety}}$ (right). In both cases gravity leads to
    asymptotically free behavior of the scalar self-coupling. 
    Negative $\lambda$ lead to instability and thus excluded.}
  \label{neg}
\end{figure}

\begin{figure}
  \includegraphics[width=\textwidth]{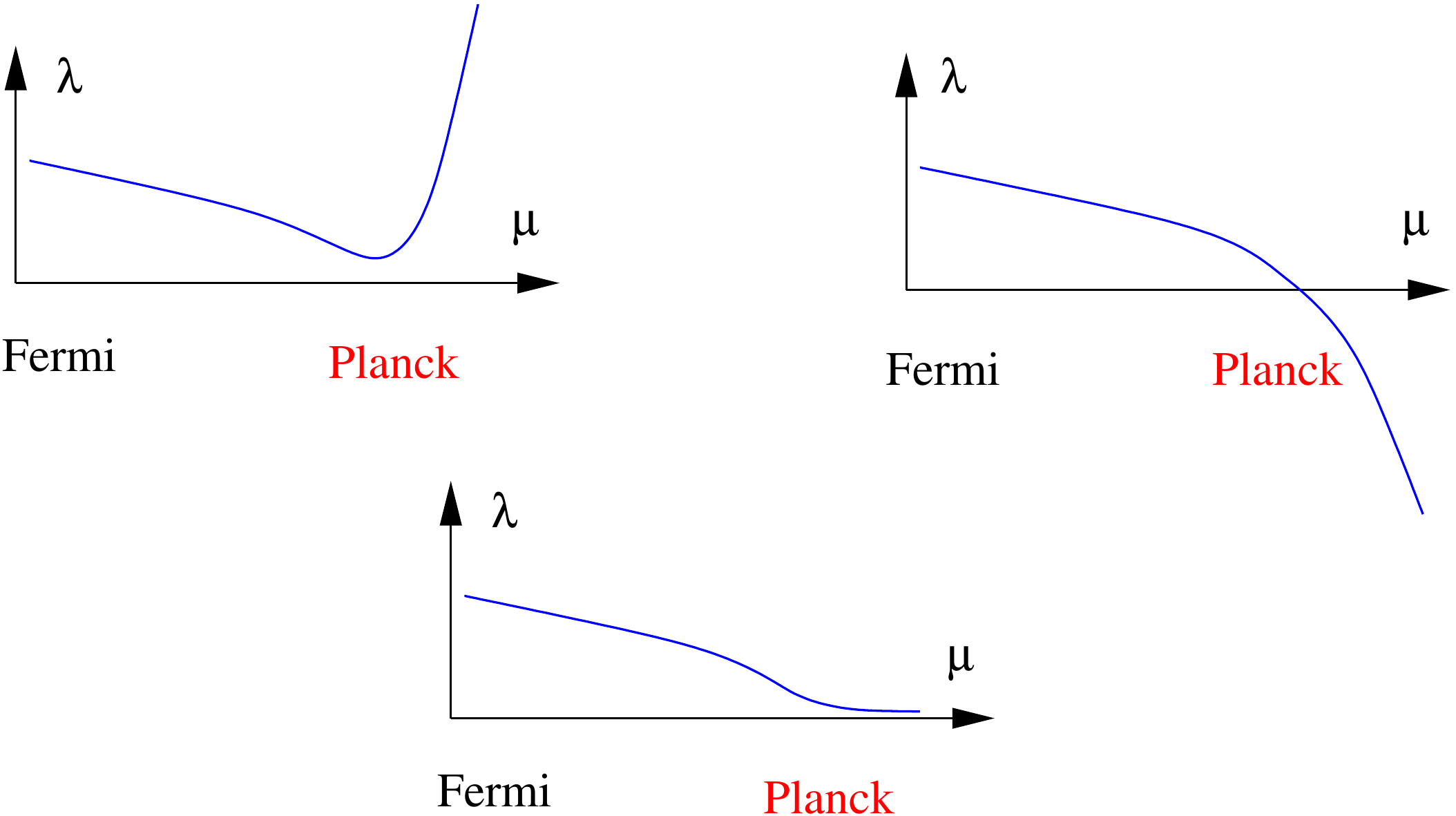}
  \caption{Schematic depiction of the behavior of the scalar
    self-coupling if $a_h>0$ for $M_H > M_{\min}^{\text{safety}}$,
    leading to Landau-pole behavior (left), $M_H>
    M_{\min}^{\text{safety}}$, leading to instability (right) and
    $M_H= M_{\min}^{\text{safety}}$, asymptotically safe behavior
    (middle). Only this choice is admissible. }
  \label{pos}
\end{figure}

The prediction of the Higgs boson mass from the requirement of
asymptotic safety of the SM is found as follows
\cite{Shaposhnikov:2009pv}.  Consider the SM running of the coupling
constants and add to the $\beta$-functions extra terms coming from
gravity, deriving their structure from dimensional analysis:
\be
  \label{betagrav}
  \beta_h^{\text{grav}} = \frac{a_h}{8\pi} \frac{\mu^2}{M_P^2(\mu)} h,
\ee
where $a_1$, $a_2$, $a_3$, $a_y$, and $a_\lambda$ are some constants
(anomalous dimensions) corresponding to the gauge couplings of the SM
$g$, $g'$, $g_s$, the top Yukawa coupling $y_t$, and the Higgs
self-coupling $\lambda$.  In addition,
\be
  \label{runMP}
  M_P^2(\mu)\simeq M_P^2 +2 \xi_0 \mu^2
\ee
is the running Planck mass with $\xi_0\approx 0.024$ following from
numerical solutions of functional RG equations
\cite{Reuter:1996cp,Percacci:2003jz,Narain:2009fy}.  Now, require that
the solution for all coupling constants is finite for all $\mu$ and
that $\lambda$ is always positive.  The SM can only be asymptotically
safe if $a_1$, $a_2$, $a_3$, $a_y$ are all negative, leading to
asymptotically safe behavior of the gauge and Yukawa couplings.  For
$a_\lambda<0$ we are getting the interval of admissible Higgs boson
masses, $M_{\min}^{\text{safety}} < M_H < M_{\max}^{\text{safety}}$. 
However, if $a_\lambda>0$, as follows from computations of
\cite{Percacci:2003jz,Narain:2009fy}, only one value of the Higgs
boson mass $M_H=M_{\min}^{\text{safety}}$ leads to asymptotically safe
behavior of $\lambda$.  As is explained in
\cite{Shaposhnikov:2009pv}, this behavior is only possible provided
$\lambda(M_P)\approx0$ and $\beta_\lambda(\lambda(M_P))\approx0$. 
And, due to miraculous coincidence of $\mu_0$ and $M_P$, the
difference $\Delta{}m^{\text{safety}} \equiv
M_{\min}^{\text{safety}}-M_{\min}$ is extremely small, of the order
$\unit[0.1]{GeV}$.  The evolution of the Higgs self-coupling for the
case of $a_h<0$ is shown in Fig.~\ref{neg}, and for the case
$a_h>0$ in Fig.~\ref{pos}.

In fact, in the discussion of the asymptotic safety of the SM one can
consider a more general situation, replacing the Planck mass in
eq.~(\ref{runMP}) by some cutoff scale $\Lambda = \kappa M_P$. 
Indeed, if the Higgs field has non-minimal coupling with gravity (see
below), the behavior of the SM coupling may start to change at
energies smaller than $M_P$ by a factor $1/\xi$, leading to an
expectation for the range of $\kappa$ as $1/\xi \lsim \kappa \lsim
1$.  Still, the difference between $M_{\min}$ and
$M_{\min}^{\text{safety}}$ remains small even for $\kappa \sim
10^{-4}$, where $M_{\min}^{\text{safety}}\simeq\unit[128.4]{GeV}$,
making the prediction $M_H \simeq M_{\min}$ sufficiently stable
against specific details of Planck physics within the asymptotic
safety scenario.

%%%%%%%%%%%%%%%%%%%%%%%%%%%%%%%%%%%%%%%%%%%%%%%%%%%%%%%%%%%%%%%%%%%%%
%%%%%%%%%%%%%%%%%%%%%%%%%%%%%%%%%%%%%%%%%%%%%%%%%%%%%%%%%%%%%%%%%%%%%%

%%%%%%%%%%%%%%%%%%%%%%%%%%%%%%%%%%%%%%%%%%%%%%%%%%%%%%%%%%%%%%%%%%%%%%
\subsection{$M_{\min}$ and cosmology}
\label{sec:cosmo}

It is important to note that if the mass of the Higgs boson is smaller
than the stability bound $M_{\min}$, this does not invalidate the SM. 
Indeed, if the life-time of the metastable SM vacuum exceeds the age
of the Universe (this is the case when $M_H > M_{\text{meta}}$, with
$M_{\text{meta}} \simeq \unit[111]{GeV}$~\cite{Espinosa:2007qp}) then
finding a Higgs boson in the mass interval $M_{\text{meta}}< M_H <
M_{\min}$ would simply mean that we live in the metastable state with
a very long lifetime. Of course, if the Higgs boson were discovered
with a mass below $M_{\text{meta}}$, this would prove that there
\emph{must be} new physics between the Fermi and Planck scales,
stabilizing the SM vacuum state. However, the latest LEP results,
confirmed recently by LHC, tell us that in fact $M_H>M_{\text{meta}}$,
and, therefore, that the presence of a new energy scale is not
required, if only the metastability argument is used.

The bound $M_H>M_{\text{meta}}$ can be strengthened if thermal
cosmological evolution is considered~\cite{Espinosa:2007qp}.  After
inflation the universe should find itself in the vicinity of the SM
vacuum and stay there till present.  As the probability of the vacuum
decay is temperature dependent, the improved Higgs boson mass bound is
controlled by the reheating temperature after inflation (or maximal
temperature of the Big Bang). The latter is model dependent, leading
to the impossibility to get a robust bound much better than
$M_{\text{meta}}$.  For example, in $R^2$ inflation
\cite{Starobinsky:1980te,Gorbunov:2010bn} the reheating temperature is
rather low, $T\sim\unit[10^9]{GeV}$ \cite{Gorbunov:2010bn}, leading to
the lower bound $\unit[116]{GeV}$~\cite{Bezrukov:2011gp} on the Higgs
boson mass, which exceeds $M_{\text{meta}}$ only by $\unit[4]{GeV}$.

However, if no new degrees of freedom besides those already present in
the SM are introduced and the Higgs boson plays the role of inflaton,
the bound $M_H \gsim M_{\min}$ reappears, as is discussed below.

%%%%%%%%%%%%%%%%%%%%%%%%%%%%%%%%%%%%%%%%%%%%%%%%%%%%%%%%%%%%%%%%%
\subsection{Higgs inflation}
\label{sec:HI}

\begin{figure}
  \includegraphics[width=\textwidth]{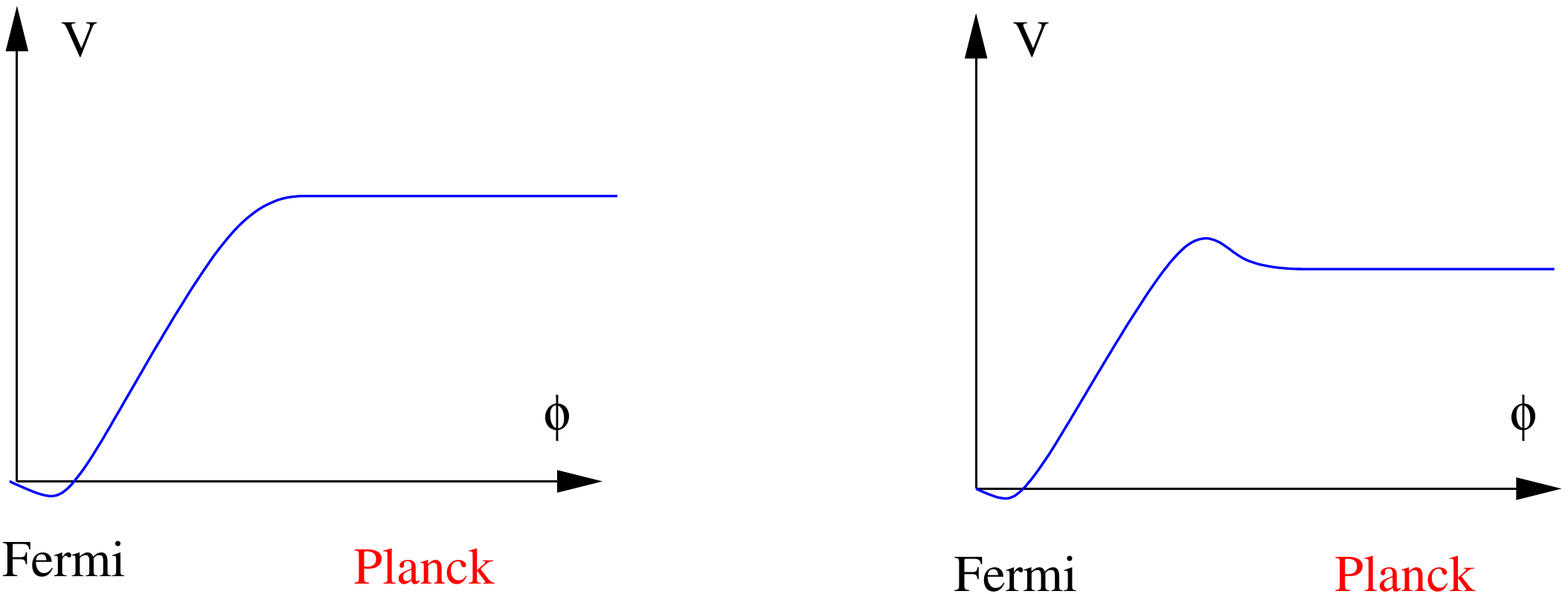}
  \caption{Schematic depiction of the effective potential $V$ for the Higgs
    field in the Higgs-inflationary theory in the Einstein frame for
    $M_H>m_{\min}^{\text{inflation}}$ (left) and
    $M_H<m_{\min}^{\text{inflation}}$ (right).}
  \label{infl}
\end{figure}

The inclusion of a non-minimal interaction of the Higgs field with
gravity, given by the Lagrangian $\xi |\phi|^2 R$, where $R$ is the Ricci
scalar, changes drastically the behavior of the Higgs potential in the
region of large Higgs fields $\phi> M_{\text{inflation}}\simeq
M_P/\sqrt{\xi}$ \cite{Bezrukov:2007ep}.  Basically, the potential
becomes flat at $\phi> M_{\text{inflation}}$, keeping the value it
acquired at $\phi\simeq M_P/\sqrt{\xi}$. This feature leads to a possibility
of Higgs-inflation: if the parameter $\xi$ is sufficiently large, $700
< \xi < 10^5$, \cite{Bezrukov:2009db} the Higgs boson of the SM can
make the Universe flat, homogeneous and isotropic, and can produce the
necessary spectrum of primordial fluctuations.  The possibility of the
Higgs inflation is also strongly related to the value of the Higgs
boson mass: the successful inflation can only occur if
$M_{\min}^{\text{inflation}} < M_H < M_{\max}^{\text{inflation}}$.  The
upper limit $M_{\max}^{\text{inflation}}$ comes from the requirement of
the validity of the SM up to the inflation scale
$M_{\text{inflation}}$.  Near $M_{\min}^{\text{inflation}}$ the
behavior of the effective potential in the Einstein frame changes as
shown in Fig.~\ref{infl}: if $M_H<M_{\min}^{\text{inflation}}$ the
``bump'' in the Higgs potential prevents the system to go to the SM
vacuum state. As in the previous case, these bounds can be formulated
with the use of the Higgs self-coupling $\lambda$.  Basically, it must
be perturbative and positive for all energy scales below
$M_{\text{inflation}}$. Though any Higgs boson mass in the interval
$M_{\min}^{\text{inflation}} < M_H < M_{\max}^{\text{inflation}}$ can
lead to successful inflation, the value  $M_{\min}^{\text{inflation}}$
is somewhat special. For the lower part of the admitted interval the
value of the non-minimal coupling $\xi$ reaches its minimal value
$\xi\simeq 700$, extending the region of applicability of perturbation
theory~\cite{Bezrukov:2009db,Bezrukov:2010jz,Barvinsky:2009fy}.

%%%%%%%%%%%%%%%%%%%%%%%%%%%%%%%%%%%%%%%%%%%%%%%%%%%%%%%%%%%%%%%%%%%%%
%%%%%%%%%%%%%%%%%%%%%%%%%%%%%%%%%%%%%%%%%%%%%%%%%%%%%%%%%%%%%%%%%%%%%%%%

The computation of the lower bound on the Higgs boson mass from
inflation is more complicated. It is described in detail in
\cite{Bezrukov:2008ej,Bezrukov:2009db}.  Basically, one has to compute
the Higgs potential in the chiral electroweak theory associated with
large values of the Higgs field and find when the slow-roll
inflation in this potential can give the large-scale perturbations
observed by the COBE satellite. The outcome of these computations,
however, can be formulated in quite simple terms: for inflationary
bound find $M_{\min}^{\text{inflation}}$ from the condition
$\lambda(\mu)>0$ for all $\mu < M_{\text{inflation}}$
\cite{Bezrukov:2009db}.  A priori, the inflationary bound could have
been very different from $M_{\min}$ and thus from
$M_{\min}^{\text{stability}}$. Indeed, both $M_{\min}$ and
$M_{\min}^{\text{stability}}$ know \emph{nothing} about the Planck
scale and are defined entirely within the SM, whereas the inflationary
bound does use $M_P$.  However, the remarkable numerical coincidence,
between $\mu_0$ and $M_P$, makes $M_{\min}$ and
$M_{\min}^{\text{inflation}}$ practically the same.  The coupling
constant  $\lambda$ evolves very slowly near the Planck scale, so that
the regions for the Higgs boson mass following from the conditions
$\lambda(\mu)>0$ for $\mu < M_P$ and $\mu < M_{\text{inflation}}$ are
almost identical.  This leads to the result that $\Delta
m^{\text{inflation}}\equiv M_{\min}^{\text{inflation}} -M_{\min}
\simeq\unit[-0.1-0.2]{GeV}$. This number is derived within the SM
without addition of any higher dimensional operators. 

One must note, that simple scale analysis leads to the unitarity
violation in the Higgs inflation below the Planck energy scale
\cite{Burgess:2009ea,Barbon:2009ya,Burgess:2010zq,Hertzberg:2010dc}.
This means, that calculations in the Higgs inflationary models should
be done with some additional assumptions about the high energy
physics, formulated in \cite{Bezrukov:2010jz}, specifically the
approximate scale invariance at high field backgrounds.  As is
explained in \cite{Bezrukov:2010jz}, adding to the SM
higher-dimensional operators with a Higgs-field dependent cutoff
modifies the lower bound on the Higgs boson mass in Higgs inflation.
If these operators are coming with ``natural'' power counting
coefficients (for exact definition see \cite{Bezrukov:2010jz}) the
sensitivity of the Higgs boson mass bound to unknown details of
ultraviolet physics is rather small $\Delta
M_{\min}^{\text{inflation}}\simeq\unit[0.6]{GeV}$
\cite{Bezrukov:2010jz}.  At the same time, it is certainly not
excluded that the change of $M_{\min}^{\text{inflation}}$ can be
larger.

\section{Conclusions}
\label{sec:concl}

If the SM Higgs boson will be discovered at LHC in the remaining mass
interval  $115.5 < M_H < \unit[127]{GeV}$ not excluded at 95\% 
\cite{ATLAS:2012ae,Chatrchyan:2012tx}, there is no necessity for a new
energy scale between the Fermi and Planck scales. The EW theory
remains in a weakly coupled region all the way up to $M_P$, whereas
the SM vacuum state lives longer than the age of the Universe. If the
SM Higgs boson mass will be found to \emph{coincide} with $M_{\min}$
given by (\ref{eq:mmin}), this  would  put a strong argument in favor
of the \emph{absence} of such a scale and indicate that the
electroweak symmetry breaking may be associated with the physics at
the Planck scale.

The experimental precision in the Higgs boson mass measurements at the
LHC can eventually reach $\unit[200]{MeV}$ and thus be much smaller
than the present theoretical ($\sim\unit[1\text{--}2]{GeV}$) and
experimental ($\sim\unit[5]{GeV}$, $2\sigma$) uncertainties in
determination of $M_{\min}$. The largest uncertainty comes from the
measurement of the mass of the top quark. It does not look likely that
the LHC will substantially reduce the error in the top quark mass
determination.  Therefore, to clarify the relation between the Fermi
and Planck scales a construction of an electron-positron or muon
collider with a center-of-mass energy of $\sim\unit[200+200]{GeV}$
(Higgs and t-quark factory) would be needed.  This would be decisive
for setting up the question about the necessity for a new energy scale
besides the two ones already known---the Fermi and the Planck scales.
In addition, this will allow to study in detail the properties of the
two heaviest particles of the Standard Model, potentially most
sensitive to any types on new physics.

Surely, even if the SM is a valid effective field theory all the way
up the the Planck scale, it cannot be complete as it contradicts to a
number of observations.  We would like to use this opportunity to
underline once more that the confirmed observational signals in favor
of physics beyond the Standard Model which were not discussed in this
paper (neutrino masses and oscillations, dark matter and baryon
asymmetry of the Universe) can be associated with new physics
\emph{below} the electroweak scale, for reviews see
\cite{Shaposhnikov:2007nj,Boyarsky:2009ix} and references
therein.\footnote{As for the dark energy, it may be related to a
massless dilaton realizing spontaneously broken scale invariance
\cite{Shaposhnikov:2008xb,Shaposhnikov:2008xi}.}  The minimal
model---$\nu$MSM, contains, in addition to the SM particles, three
relatively light singlet Majorana fermions. These fermions could be
responsible for neutrino masses, dark matter and baryon asymmetry of
the Universe.  The $\nu$MSM predicts that the LHC will continue to
confirm the Standard Model and see no deviations from it. At the same
time, new experiments at the high-intensity frontier, discussed in
\cite{Gorbunov:2007ak}, may be needed to uncover the new physics below
the Fermi scale. In addition, new observations in astrophysics,
discussed in \cite{Boyarsky:2009ix}, may shed light to the nature of
Dark Matter. As the running of couplings in the $\nu$MSM coincides
with that in the SM, all results of the present paper are equally
applicable to the $\nu$MSM.

\subsection*{Acknowledgements}

The work of M.S. has been supported by the Swiss National Science
Foundation. The work of M.Yu.K. and B.A.K. was supported in part by
the German Federal Ministry for Education and Research BMBF through
Grants No.\ 05~HT6GUA, 05~HT4GUA/4 by the German Research Foundation
DFG through the Collaborative Research Centre No.~676 \emph{Particles,
Strings and the Early Universe---The Structure of Matter and Space
Time,} and by the Helmholtz Association HGF through the Helmholtz
Alliance Ha~101 \emph{Physics at the Terascale.}  We thank A.
Boyarsky, K. Chetyrkin, D. Gorbunov, F.~Jegerlehner, G.~Passarino,
O.~Ruchayskiy, and M. Zoller for helpful discussions, collaboration
and interest to our work. M.Yu.K. is indebted to Fred Jegerlehner for
a fruitful long time collaboration on developing and study \MSb\
scheme beyond one-loop order in framework of quantum field models with
spontaneously symmetry breaking, and in particular for collaboration
on \cite{Jegerlehner:2003sp}, the results and methods of which were
heavily used in the present work.

\subsection{Note added}

After our paper was submitted to the electronic preprint archive arXiv
(on May 13) a number of events happened, which require its update.
First, extra corrections to the matching procedure at low energy
scale, not computed in our work, were found in \cite{Degrassi:2012ry}
(May 29).\footnote{Preliminary results of Ref.~\cite{Degrassi:2012ry}
  were presented in Ref.~\cite{Strumia:Higgstalk2012}.  However, the
  $O(\alpha \alpha_s)$ threshold correction to the Higgs self-coupling
  is not specified in Ref.~\cite{Strumia:Higgstalk2012}, but is
  presented here for the first time.  Moreover, the conclusions
  reached in Ref.~\cite{Strumia:Higgstalk2012} disagree with ours.}
Ref.~\cite{Degrassi:2012ry} finds agreement with our results on the
$\alpha\alpha_s$ order in the Higgs self-coupling constant.  In
addition, this paper computed a part of the $O(\alpha^2)$ corrections
to the top Yukawa and Higgs coupling constants in the ``gauge-less''
limit of the Standard model, i.e.\ the two-loop terms containing the
top Yukawa and scalar self-coupling were accounted for.  The overall
effect of these terms happened to be quite small.  The corrections
shift the benchmark Higgs mass up by \unit[0.2]{GeV}, and reduce the
sensitivity of the results to the normalisation point from
\unit[1.2]{GeV} to \unit[0.8]{GeV}, decreasing somewhat the
theoretical error-bars, see Fig.~\ref{fig:mu0dep}.

Second, the discovery of the Higgs-like resonance was announced at
CERN by ATLAS and CMS collaborations \cite{ATLAS:2012gk,CMS:2012gu}.
According to CMS,
\be  
  M_H = \unit[125.3 \pm  0.4 \text{(stat)} \pm 0.5 \text{(syst)}]{GeV},
\ee
while ATLAS gives a slightly higher value
\be
  M_H = \unit[126 \pm  0.4 \text{(stat)} \pm 0.4 \text{(syst)}]{GeV}.
\ee

And, finally, an updated results on the mass of the top quark were
announced at ICHEP 2012 (July 9), see \cite{TopICHEP2012,Lancaster:2011wr}.
The combination of the Tevatron results reads:
\be
  m_t=\unit[173.2 \pm 0.6 \text{(stat)}\pm 0.8 \text{(syst)}]{GeV}
     =\unit[173.2\pm 1.0]{GeV},
\ee
whereas the present LHC value is
\be
  m_t=\unit[173.3 \pm 0.5 \text{(stat)}\pm 1.3 \text{(syst)}]{GeV}
     =\unit[173.3\pm 1.4]{GeV}.
\ee
The central values of the top mass are somewhat higher (by
\unit[0.3--0.4]{GeV}) than those which were given by the Particle Data
Group at the time we were writing our paper.

\begin{figure}
  \centering
  \includegraphics{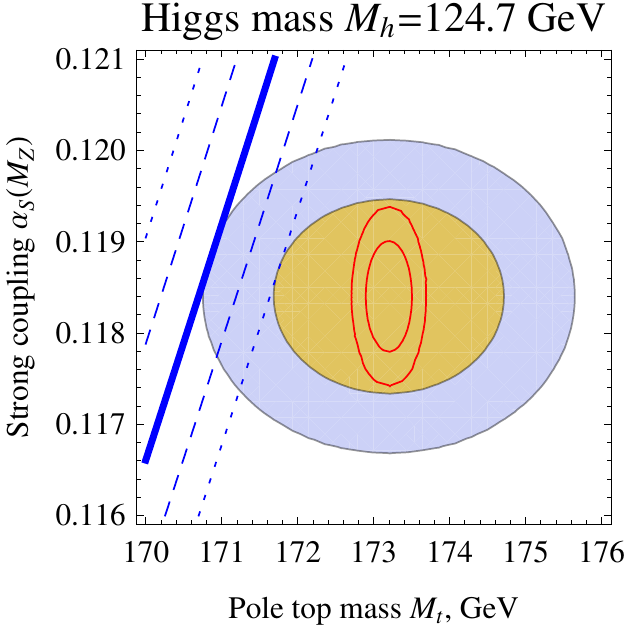}
  \includegraphics{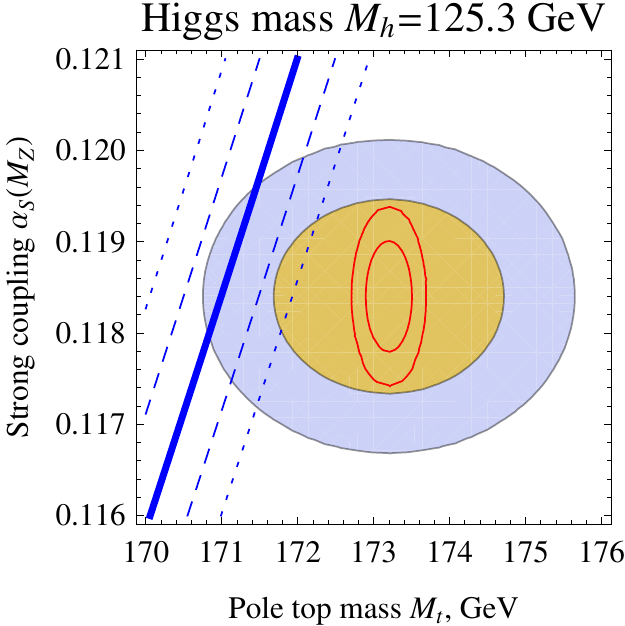}\\
  \includegraphics{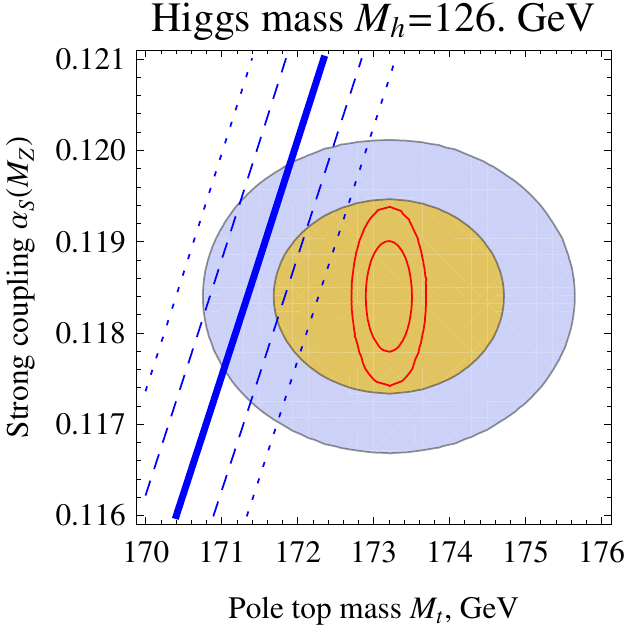}
  \includegraphics{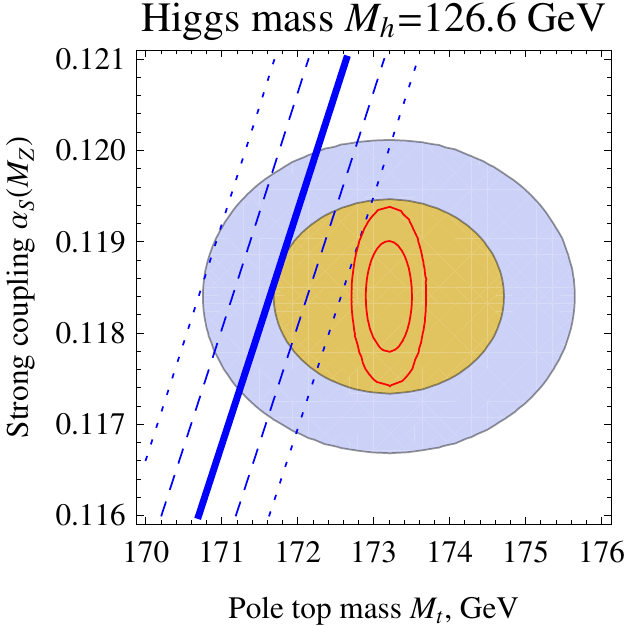}
  \caption{Update of Fig.~\ref{fig:asmt}.  The values of the strong
    coupling constant $\aS$ and top mass $M_t$ corresponding to a
    minimal Higgs boson mass $M_{\min}$ coinciding with the recent CMS
    and ATLAS values \unit[125.3]{GeV} and \unit[126]{GeV}, and to the
    Higgs mass values shifted by approximately
    $1\sigma\sim\unit[0.6]{GeV}$.  The latest Tevatron
    68\% and 95\% experimentally allowed regions for $\aS$ and $M_t$
    are given by shaded areas.  The dashed (dotted) lines correspond
    to $\unit[1]{GeV}$ ($\unit[2]{GeV}$) uncertainty in the $M_{\min}$
    theoretical determination.  The red lines in the center correspond
    to the expected precision from a $e^+e^-$ collider.}
  \label{fig:newasmt}
\end{figure}

In Fig.~\ref{fig:newasmt}, which is an update of Fig.~\ref{fig:asmt},
we show the changes due to the experimental shift of the top mass (we
take the Tevatron result as having smaller errors) and due to the additional
two-loop corrections found in \cite{Degrassi:2012ry} for the ATLAS and
CMS values of the Higgs mass.  The value of the benchmark
Higgs mass is inside $2\sigma$ contours on the $m_t,\alpha_s$ plane
for the Atlas and CMS values of $M_H$, entering the $1\sigma$ contours
if theoretical uncertainties are incorporated.  The red curves show
the shrinking the size of uncertainties in $m_t$ and $\alpha_s$ (1 and
$2\sigma$) which can be achieved with an $e^+e^-$ collider operated as
a in $t\bar t$ factory on a lepton
collider (this estimate is taken from
\cite{Accomando:1997wt,*AguilarSaavedra:2001rg,*Abe:2001nnb,
  *Abe:2001npb,*Abe:2001nqa,*Abe:2001nra,*Abe:2001gc,*Djouadi:2007ik}
and \cite{Winter_ILC:2001} discussing the ILC physics).

It is important also to note, that it is difficult to determine which
renormalization scheme corresponds to the numerical value of $m_t$
quoted by experiments.  The $m_t$ determinations by the Tevatron and
LHC collaborations \cite{TopICHEP2012,Blyweert:2012bq} are based on
Monte Carlo event generators implemented with LO hard-scattering
matrix elements.  Although it is plausible and likely that the
experimental $m_t$ values thus extracted are close to the pole mass of
the top quark, this is by no means guaranteed with the due theoretical
rigor \cite{Schwanenberger:ICHEP2012}.  In fact, rigorous
determinations of the top-quark pole mass from total production cross
sections yield somewhat smaller values, albeit with larger errors
\cite{Alekhin:2012py,Hoang:2009yr,Hoang:2008yj}.  For the time being,
we adopt the working hypothesis that the experimental values of $m_t$
\cite{Blyweert:2012bq,TopICHEP2012} correspond to the pole mass,
bearing in mind that this is probably subject to change once a proper
NLO treatment of the resonating top-quark propagators is implemented.

Thus, all our considerations remain in force and call for further
improvement of the theoretical computations, which should account for
all $O(\alpha^2)$ corrections to the mapping procedure, and for a
construction of a $t\bar t$ factory to pin down $\alpha_s$ and $m_t$.  A
recent paper \cite{Alekhin:2012py}, where uncertainties in
determination $m_t$ and $\alpha_s$ have been analyzed, reached exactly
the same conclusion.

%%%%%%%%%%%%%%%%%%%%%%%%%%%%%%%%%%%%%%%%%%%%%%%%%%%%%%%%%%%%%%%%%%%%%%%%
\appendix

\newcommand{\half}{\frac{1}{2}}
\newcommand{\amu}{\alpha(\mu)}
\newcommand{\amz}{\alpha(M_Z)}
\newcommand{\sswmz}{\sin^2\theta(M_Z)}
\newcommand{\sswOS}{s_W^2}
\renewcommand{\aS}{{\alpha_s}}

\newcommand{\ep}{\varepsilon}
\newcommand{\Gl}[2]{{\mbox{Gl}}_{#1}\left(#2\right)}
\newcommand{\Li}[2]{{\mbox{Li}}_{#1}\left(#2\right)}
\newcommand{\Cl}[2]{{\mbox{Cl}}_{#1}\left(#2\right)}
\newcommand{\Ls}[2]{{\mbox{Ls}}_{#1}\left(#2\right)}
\newcommand{\LS}[3]{{\mbox{Ls}}_{#1}^{(#2)}\left(#3\right)}
\newcommand{\Lsc}[2]{{\mbox{Lsc}}_{#1\!}\left(#2\right)}
\newcommand{\ST}[2]{\left[ #1 \atop #2\right]}
\newcommand{\Snp}[2]{{\mbox{S}}_{#1\!}\left(#2\right)}
\newcommand{\HA}[2]{{\mbox{H}}^{(#1)}_{#2}}
\newcommand{\MIN}{\mbox{i}^{1-n}}
\newcommand{\MSbn}{$\overline{\rm MS}$}

\newcommand{\Z}{\Delta}

\newcommand{\SIGMA}{\sigma_{\alpha \alpha_s}}

\renewcommand{\theequation}{\thesection.\arabic{equation}}
%\renewcommand{\thefootnote}{\fnsymbol{footnote}}

%%%%%%%%%%%%%%%%%%%%%%%%%%%%%%%%%%%%%%%%%%%%%%%%%%%%%%%%%%%%%%%%%%
\section{$O(\alpha\aS)$ electroweak corrections to the
  top Yukawa and Higgs self couplings in Standard Model}
\label{app:PoleMatching}

The evaluation of radiative corrections to the relations between \MSb\ parameters (coupling constants) 
and masses of particles includes two steps: evaluation of radiative corrections between 
the Fermi constant $G_F$ and its \MSb\ counterpart \cite{Sirlin:1980nh} 
%which we denote as $v(\mu^2)$  
(see \cite{Marciano:1980pb,Burgers:1989bh,Actis:2006rc} for recent reviews)  
and the evaluation of the radiative corrections between \MSb\ and pole masses. 

  The one-loop electroweak corrections ${\cal O}(\alpha)$ to the relation
between the self-coupling $\lambda(\mu^2)$ and the pole mass of the Higgs boson 
was obtained in \cite{Sirlin:1985ux} and to the  relation between 
the Yukawa coupling $y_t$ and the pole mass of top quark was found in~\cite{Hempfling:1994ar}.  
The corresponding ingredients for the 2-loop mixed electroweak-QCD corrections
were evaluated in 
\cite{Chang:1981qq,Djouadi:1987di,Kniehl:1989yc,Djouadi:1994gf,Jegerlehner:2003py,Jegerlehner:2003sp},
but has never been assembled. 
We performed independent (re)calculations of all $O(\alpha)$ and $O(\alpha \alpha_s)$ contributions. 
In the following we will denote the on-shell masses by capital $M$ and the \MSb\ masses by lowercase $m$. 

%%%%%%%%%%%%%%%%%%%%%%%%%%%%%%%%%%%%%%%%%%%%%%%%%%%%%%%%%%%%%%%%%%%%%%%%%%%%%%%%%%%%%%%%%%%%%%%%%%%%%%%
%%%%%%%%%%%%%%%%%%%%%%%%%%%%%%%%%%%%%%%%%%%%%%%%%%%%%%%%%%%%%%%%%%%%%%%%%%%%%%%%%%%%%%%%%%%%%%%%%%%%%%%
%%%%%%%%%%%%%%%%%%%%%%%%%%%%%%%%%%%%%%%%%%%%%%%%%%%%%%%%%%%%%%%%%%%%%%%%%%%%%%%%%%%%%%%%%%%%%%%%%%%%%%%
\subsection{$O(\alpha \alpha_s)$ corrections to the relation between on-shell and \MSb\ Fermi constant }
The relation between the Fermi coupling constant and the bare parameters is as follows \cite{Sirlin:1980nh}: 
\begin{eqnarray}
  \frac{G_F}{\sqrt{2}} & = & \frac{g_0^2}{8 m_{W,0}^2} \left\{ 1 + \Delta R_0 \right\} ,
\end{eqnarray}
where $\Delta R_0$ includes unrenormalized electroweak corrections
and $g_0,m^2_{W,o}$  are the SU(2) coupling constant and the bare $W$ boson mass
(see for details \cite{Marciano:1980pb,Burgers:1989bh,Actis:2006rc}).
After performing \MSb\ renormalization this relation has the following form:
\begin{eqnarray}
  \frac{G_F}{\sqrt{2}}
  = \frac{G_F(\mu^2)}{\sqrt{2}} 
  (1 + \Delta_{G_F, \alpha}  + \Delta_{G_F, \alpha \alpha_s}  + \cdots ) .
  \label{G}
\end{eqnarray}
where on the r.h.s.\ all masses and coupling constants are taken in the \MSb\
renormalization scheme.
%(depend on value of renormalized scale $\mu$ and satisfy to the renormalization group equation). 
The one-loop coefficient, $\Delta_{G_F, \alpha}$, is known from
\cite{Sirlin:1980nh} and for $N_c=3$, $C_F = 4/3$ and $m_b=0$ has the following
form:
\begin{eqnarray}
  \Delta_{G_F,\alpha}  & = & 
  \frac{g^2}{16 \pi^2}
  \Biggl\{
  \frac{m_t^4}{m_W^2 m_H^2} \left( 6 - 6 \ln \frac{m_t^2}{\mu^2} \right)
  + \frac{m_t^2}{m_W^2}    \left( -\frac{3}{4} + \frac{3}{2} \ln \frac{m_t^2}{\mu^2} \right)
  + \frac{m_H^2}{m_W^2}    \left( - \frac{7}{8} + \frac{3}{4} \ln \frac{m_H^2}{\mu^2} \right)
  \nonumber \\ && 
  + \frac{m_Z^4}{m_H^2 m_W^2}    \left( -\frac{1}{2} + \frac{3}{2} \ln \frac{m_Z^2}{\mu^2} \right)
  + \frac{m_W^2}{m_H^2}    \left( - 1 + 3 \ln \frac{m_W^2}{\mu^2} \right)
  - \frac{3}{4} \frac{m_W^2}{m_H^2-m_W^2 }   \ln \left( \frac{m_W^2}{m_H^2} \right)
  \nonumber \\ && 
  + \frac{m_Z^2}{m_W^2}    \left( \frac{5}{8} + \frac{17}{4} \ln \frac{m_W^2}{\mu^2}  - 5 \ln \frac{m_Z^2}{\mu^2} \right)
  - \frac{3}{4} \ln \frac{m_W^2}{\mu^2}
  - \frac{3}{4} \ln \frac{m_H^2}{\mu^2}
  \nonumber \\ && 
  - \frac{17}{4}\frac{m_Z^2}{m_W^2 \sin^2 \theta_W}   \ln \left( \frac{m_W^2}{m_Z^2} \right)
  + \frac{5}{4}  
  + \frac{7}{2 \sin^2 \theta_W}  \ln \left( \frac{m_W^2}{m_Z^2} \right)
  \Biggr\} 
  ,
\end{eqnarray}
Here, $\sin^2 \theta_W$ is defined in the \MSb\ scheme as
\begin{equation}
\sin^2 \theta_W 
\equiv 
\sin^2 \theta^{\overline{MS}}_W(\mu^2) 
= \frac{g'^2(\mu^2)}{g^2(\mu^2)+g'^2(\mu^2)}
= 1 - \frac{m_W^2(\mu^2)}{m_Z^2(\mu^2)} ,
\label{sin}
\end{equation}
where $g'(\mu^2)$ and $g(\mu^2)$ are the U(1) and SU(2) \MSb\ gauge coupling constants, respectively. 
The matching conditions between the \MSb\ parameter, defined by Eq.~(\ref{sin}), and its on-shell version, \cite{Sirlin:1980nh}, follows 
from identification 
\begin{equation}
  \label{swOSdef}
  \sin^2 \theta_W^{OS} = 1 - \frac{M_W^2}{M_Z^2},   
\end{equation}
where $M_Z$ and $M_W$ are the pole masses of the gauge bosons (see detailed discussion in \cite{Jegerlehner:2001fb,Jegerlehner:2002er,Jegerlehner:2002em}). 
The evaluation of the mixed QCD-EW coefficient, $\Delta_{G_F, \alpha \alpha_s}$,
is reduced to the evaluation of the $O(\alpha \alpha_s)$ corrections 
to the $W$ boson self-energy at zero momenta transfer and may be written in the following way \cite{Halzen:1990je,Fanchiotti:1992tu,Kniehl:1993jc,Kniehl:1994ph,Djouadi:1993ss}:
\begin{align}
\Delta_{G_F, \alpha \alpha_s} 
&\equiv 
2 g_R^2 Z_{g,\alpha \alpha_s}
- 
\left[ 
Z_{W,\alpha \alpha_s}
- 
Z_{m_t^2,\alpha_s} m_t^2 \frac{\partial}{\partial m_t^2}\frac{\Pi_{WW,\alpha}(0)}{m_W^2(\mu^2)} 
- 
\frac{\Pi_{WW,\alpha \alpha_s}(0)}{m_W^2(\mu^2)} 
\right] 
\label{G-alpha-alpha-s}
\\ 
&= 
%\SIGMA
C_f N_c \frac{g^2 g_s^2}{ (16 \pi^2)^2 }
\frac{m_t^2}{m_W^2}  
\Biggl[ 
20 \frac{m_t^2}{m_H^2}
-\frac{13}{8} 
+ \zeta_2  
\nonumber\\ &
\hphantom{=C_f N_c \frac{g^2 g_s^2}{ (16 \pi^2)^2 }\frac{m_t^2}{m_W^2}\Biggl[ }
+
\left( 1 \!-\! 20 \frac{m_t^2}{m_H^2} \right) \ln \left( \frac{m_t^2}{\mu^2} \right)
- \left( \frac{3}{2} \!-\! 12 \frac{m_t^2}{m_H^2} \right) \ln^2 \left( \frac{m_t^2}{\mu^2} \right) 
\Biggr] ,
\nonumber
\end{align}
where 
% $$
% \SIGMA \equiv C_f N_c \frac{g^2 g_s^2}{ (16 \pi^2)^2 }
% $$
for 
$ 
Z_{g,\alpha \alpha_s},
Z_{g,\alpha \alpha_s}$ and $Z_{W,\alpha \alpha_s} $ we used the results\footnote{There are typos in Eq.~(4.41) of \cite{Jegerlehner:2002em}: 
in all \MSb\ renormalization constants, $Z_W^{\alpha_s}$ and $Z_Z^{\alpha_s}$, 
``$m_t^2/m_W^2$'' should be replaced by ``$m_t^2/m_H^2$''
\begin{eqnarray}
Z_W^{\alpha_s} & = &
1 + \frac{g^2}{(16 \pi^2)} \frac{\alpha_s}{4 \pi} N_c C_f
\Biggl[
  \frac{1}{\ep} \left(
4 \frac{m_t^4}{m_H^2 m_W^2}  - \frac{5}{4} \frac{m_t^2}{m_W^2}
+ \frac{1}{2} n_F
                \right)
+ \frac{1}{\ep^2}
 \left( -12 \frac{m_t^4}{m_H^2 m_W^2}  + \frac{3}{2} \frac{m_t^2}{m_W^2 } \right)
\Biggr]
\nonumber \\
Z_Z^{\alpha_s} & = & 1 + \frac{g^2}{(16 \pi^2)} \frac{\alpha_s}{4 \pi} N_c C_f
\Biggl[
 \frac{1}{\ep} \left(
4 \frac{m_t^4}{m_H^2 m_W^2}  - \frac{5}{4} \frac{m_t^2}{m_W^2}
+ \frac{10}{9} n_F    \frac{m_W^2}{m_Z^2}
+ \frac{11}{18} n_F \frac{m_Z^2}{m_W^2}
- \frac{11}{9} n_F
                \right)
\nonumber \\ && \hspace{35mm}
+ \frac{1}{\ep^2}  \left( -12 \frac{m_t^4}{m_H^2 m_W^2}
+ \frac{3}{2} \frac{m_t^2}{m_W^2 } \right)
\Biggr]  ,
\nonumber 
\end{eqnarray}
}
of \cite{Jegerlehner:2002em} and $n_F$ is the number of fermion families ($n_F$
is equal to $3$ in the SM).

Using the fact, that $G_F$ is RG invariant, i.e.\ $\mu^2 \frac{d}{d \mu^2} G_F=0$,
the $\mu$-dependent terms in Eq.~(\ref{G-alpha-alpha-s})
can be evaluated explicitly from the one-loop correction and 
explicit knowledge of anomalous dimension $\gamma_{G_F}$. 
As was shown in \cite{Jegerlehner:2001fb,Jegerlehner:2002er,Jegerlehner:2002em,Jegerlehner:2003wu,Jegerlehner:2004aj}, the anomalous dimension $\gamma_{G_F}$ can be extracted
(i) via the beta-function  $\beta_\lambda$ of the scalar self-coupling and the anomalous
dimension of the mass parameter $m^2$ (in unbroken phase) 
or 
(ii) via the $\beta$-function of the SU(2) gauge coupling $g$ and the anomalous
dimension of the $W$ boson (in broken phase): 
\begin{eqnarray}
\gamma_{G_F} 
\equiv 
\mu^2 \frac{\partial}{\partial \mu^2} \ln G_F(\mu^2)  
= \frac{\beta_\lambda}{\lambda} - \gamma_{m^2} 
= 2 \frac{\beta_g}{g} - \gamma_W .
\label{RG-G}
\end{eqnarray}
Eq.~(\ref{G}) can be written as
\begin{eqnarray}
\frac{G_F}{G_F(\mu^2)} & = & 
1- \frac{g^2}{16 \pi^2} 
\left[ 
\gamma_{G_F,\alpha} L \!-\! \Delta X_{G_F,\alpha}^{(1)}
\right]
\nonumber \\ && 
+ \frac{g^2 g_s^2}{(16 \pi)^2}
\left[ 
\Delta X_{G_F,\alpha \alpha_s}^{(2)}
\!+\! 
C_{G_F,\alpha \alpha_s}^{(2,2)} L^2 
\!-\! 
C_{G_F,\alpha \alpha_s}^{(2,1)} L
\right]
,
\label{explicit-Log}
\end{eqnarray}
where $L = \ln \frac{\mu^2}{m_t^2}$ 
and the coefficients $C_{G_F,\alpha \alpha_s}^{(2,2)}$ and $C_{G_F,\alpha
  \alpha_s}^{(2,1)}$  are defined via the RG equations:  
\begin{eqnarray}
2 C_{G_F,\alpha \alpha_s}^{(2,2)} 
& = &   
Z_{m_t^2,\alpha_s} \frac{\partial}{\partial m_t^2} \gamma_{G_F,\alpha}
= 
- 6 C_f 
\left.
\left[ 
\frac{3}{2} \frac{m_t^2}{m_W^2}
- 12 \frac{m_t^4}{m_W^2 m_H^2}
\right] 
\right|_{N_c=3}
, 
\\ 
C_{G_F,\alpha \alpha_s}^{(2,1)} 
& = &  
\gamma_{G_F, \alpha \alpha_s} 
+ 
Z_{m_t^2,\alpha_s} m_t^2 \frac{\partial}{\partial m_t^2} \Delta X_{G_F,\alpha}^{(1)}
+ 
Z_{m_t^2,\alpha_s} \gamma_{G_F,\alpha} 
,
\label{G21}
\end{eqnarray}
with 
\begin{equation}
\gamma_{G_F, \alpha \alpha_s} 
= 
\left[ 
2 \frac{\beta_{g, \alpha \alpha_s}}{g}
- 2 Z_{W, \alpha \alpha_s}
\right]
= 
\frac{1}{2} N_c C_f
\left[ 
5 \frac{m_t^2}{m_W^2}
- 
16 \frac{m_t^4}{m_W^2 m_H^2} 
\right]
.
\end{equation}
Collecting all terms in Eq.~(\ref{G21})
we get
\begin{eqnarray}
\left. C_{G_F,\alpha \alpha_s}^{(2,1)} \right|_{N_c=3, C_f = \tfrac{4}{3}}
= 
4 \frac{m_t^2}{m_W^2}
- 80 \frac{m_t^4}{m_W^2 m_H^2} .
\end{eqnarray}

At the end of this section we again point out that the anomalous dimension of the vacuum expectation value $v^2(\mu^2) = 1/(\sqrt{2} G_F(\mu^2))$ 
within the diagram technique is defined by Eq.~(\ref{RG-G}) and it is not equal to the
anomalous dimension of the scalar field 
as in the effective potential approach \cite{Arason:1991ic}.
Another important property of Eq.~(\ref{RG-G}) is the appearance of an inverse power of
the coupling constant $\lambda$ 
due to the explicit inclusion of the tadpole contribution.  As consequence, the limit of zero Higgs mass, $m_H^2 =0$,
does not exist within the perturbative approach. 
The importance of the inclusion of the tadpole contribution to restore gauge invariance of on-shell counterterms was 
recognized a long time ago \cite{Fleischer:1980ub} and was explicitly included in the one-loop electroweak corrections to the matching 
conditions \cite{Sirlin:1985ux,Hempfling:1994ar}.
The RG equations 
for the mass parameters were discussed in \cite{Jegerlehner:2001fb,Jegerlehner:2002er,Jegerlehner:2002em,Jegerlehner:2003wu,Jegerlehner:2004aj}.   
%%%%%%%%%%%%%%%%%%%%%%%%%%%%%%%%%%%%%%%%%%%%%%%%%%%%%%%%%%%%%%%%%%%%%%%%%%%%%%%%%%%%%%%%%%%%%%%%%%%%%%%%%%%%%%
%%%%%%%%%%%%%%%%%%%%%%%%%%%%%%%%%%%%%%%%%%%%%%%%%%%%%%%%%%%%%%%%%%%%%%%%%%%%%%%%%%%%%%%%%%%%%%%%%%%%%%%%%%%%%%
%%%%%%%%%%%%%%%%%%%%%%%%%%%%%%%%%%%%%%%%%%%%%%%%%%%%%%%%%%%%%%%%%%%%%%%%%%%%%%%%%%%%%%%%%%%%%%%%%%%%%%%%%%%%%%
%%%%%%%%%%%%%%%%%%%%%%%%%%%%%%%%%%%%%%%%%%%%%%%%%%%%%%%%%%%%%%%%%%%%%%%%%%%%%%%%%%%%%%%%%%%%%%%%%%%%%%%%%%%%%%
%%%%%%%%%%%%%%%%%%%%%%%%%%%%%%%%%%%%%%%%%%%%%%%%%%%%%%%%%%%%%%%%%%%%%%%%%%%%%%%%%%%%%%%%%%%%%%%%%%%%%%%%%%%%%%

\subsection{$O(\alpha \aS)$ corrections to the relation between the \MSb\ and pole
  masses of the top quark}

The detailed discussion and explicit evaluation\footnote{There is typo in Eq.~(4.46) of \cite{Jegerlehner:2003py}: the common factor $C_f$ was lost. The correct result is 
$$
= \frac{\alpha_s}{4 \pi} \frac{e^2}{16 \pi^2 \sin^2 \theta_W} C_f 
\left( 
  \frac{1}{C_f} C^{(2,2)}_{\alpha \alpha_s} \ln^2 \frac{m_t^2}{\mu^2}
+ \frac{1}{C_f} C^{(2,1)}_{\alpha \alpha_s} \ln \frac{m_t^2}{\mu^2}
+ \mbox{without modifications}
\right) .
$$
However, all plots, the Eq.~(5.57) and the Maple program
\cite{MKL:pole} are correct.} have been presented in
\cite{Jegerlehner:2003py} (the results of \cite{Jegerlehner:2003py}
were also used for analysis of convergence of series representation of
the set of Feynman Diagrams evaluated in \cite{Faisst:2004gn,Eiras:2005yt}).
For our analysis is enough to write the following symbolic relation between the \MSb\
and pole masses of the top quark: 
\begin{eqnarray}
\frac{m_t(\mu^2)}{M_t} & = & 
1 
+ \sigma_\alpha
+ \sigma_{\alpha_s}
+ \sigma_{\alpha_s^2}
+ \sigma_{\alpha_s^3}
+ \sigma_{\alpha \alpha_s}
+ \cdots ,
\label{Delta-T} 
\end{eqnarray}
where $\sigma_{\alpha}$ and $\sigma_{\alpha \alpha_s}$ are defined by Eq.(5.54) or
Eq.(5.57) of \cite{Jegerlehner:2003py}.

The pure QCD corrections can be found in
\cite{Chetyrkin:1999ys,Chetyrkin:1999qi,Melnikov:2000qh} (only the value of
$\sigma_{\aS}(M_t)$ is given there, but the expression for other $\mu$ values can be readily
reconstructed from the beta functions).

%%%%%%%%%%%%%%%%%%%%%%%%%%%%%%%%%%%%%%%%%%%%%%%%%%%%%%%%%%%%%%%%%%%%%%%%%%%%%%%%%%%%%%%%%%%%%%%%%%%%%%%%%%%%%%
%%%%%%%%%%%%%%%%%%%%%%%%%%%%%%%%%%%%%%%%%%%%%%%%%%%%%%%%%%%%%%%%%%%%%%%%%%%%%%%%%%%%%%%%%%%%%%%%%%%%%%%%%%%%%%
%%%%%%%%%%%%%%%%%%%%%%%%%%%%%%%%%%%%%%%%%%%%%%%%%%%%%%%%%%%%%%%%%%%%%%%%%%%%%%%%%%%%%%%%%%%%%%%%%%%%%%%%%%%%%%
%%%%%%%%%%%%%%%%%%%%%%%%%%%%%%%%%%%%%%%%%%%%%%%%%%%%%%%%%%%%%%%%%%%%%%%%%%%%%%%%%%%%%%%%%%%%%%%%%%%%%%%%%%%%%%
%%%%%%%%%%%%%%%%%%%%%%%%%%%%%%%%%%%%%%%%%%%%%%%%%%%%%%%%%%%%%%%%%%%%%%%%%%%%%%%%%%%%%%%%%%%%%%%%%%%%%%%%%%%%%%
\subsection{$O(\alpha \alpha_s)$ corrections to the relation between the \MSb\
  and pole masses of the Higgs boson}

At the two-loop level the relation between the pole and \MSb\ masses is defined as follows:
\begin{eqnarray}
&& \hspace{-5mm}
s_{P} =  m_{0}^2 \!-\! \Pi^{(1)}_{0} 
\!-\! \Pi_{0}^{(2)} \!-\! \Pi^{(1)}_{0} \Pi_{0}^{(1)}{}'
\!-\! \Biggl[ \sum\limits_j (\delta m^2_{j,0})^{(1)} \frac{\partial}{\partial m_{j,0}^2} 
\!+\! \sum\limits_j (\delta g_{j,0})^{(1)} \frac{\partial}{\partial g_{j,0}}
\Biggr] \Pi_{0}^{(1)} 
\nonumber\\ && \hspace{-5mm}
= m^2_a 
\!-\! \left \{ \! \Pi_a^{(1)} \! \right\}_{\overline{\rm MS}}
\!-\! \left \{ \! \Pi_a^{(2)} \!+\! \Pi_a^{(1)} \Pi_a^{(1)}{}' \! \right\}_{\overline{\rm MS}},
\label{mct}
\end{eqnarray}
where the sum runs over all species of particles, $g_j = \alpha$,
$g_s$, $(\delta g_{j,0})^{(1)}$ and $(\delta m^2_{j,0})^{(1)}$ are the
one-loop counterterms for the charges and physical masses in the
\MSb\ scheme and after differentiation we put all parameters equal to
their on-shell values. The derivatives in Eq.~(\ref{mct}) correspond
to the subtraction of sub-divergences.  The genuine two-loop mass
counterterm comes from the shift of the $m_{0}^2$ term.
The relation between the bare and \MSb\ masses of the Higgs boson has the form
\begin{eqnarray}
&& 
\left( m^B_H \right)^2 
= 
\left( m^R_H(\mu^2) \right)^2 
\Biggl[ 
1 \!+\! \frac{g^2}{16 \pi^2     \ep} Z_{H,\alpha}
\nonumber \\ && 
+ \frac{g^4}{(16 \pi^2)^2} 
\left( 
\frac{1}{\ep} Z_{H,\alpha^2}^{(2,1)} 
\!+\! 
\frac{1}{\ep^2} Z_{H,\alpha^2}^{(2,2)} 
\right) 
\!+\! 
\frac{g_s^2 g^2}{(16 \pi^2)^2}  
\left( 
\frac{1}{\ep} Z_{H, \alpha \alpha_s}^{(2,1)}
\!+\! 
\frac{1}{\ep^2} Z_{H, \alpha \alpha_s}^{(2,2)} 
\right) 
\Biggr] ,
\label{bare}
\end{eqnarray}
where $g$ is the SU(2) \MSb\ renormalized coupling constant.

The exact analytical result for the $O(\alpha \alpha_s)$ two-loop quark contribution
to the Higgs-boson self-energy was calculated
in~\cite{Djouadi:1994gf,Kniehl:1993jc,Kniehl:1993jc}.
%TODO (check removal)
%(see also \cite{Kniehl:1989yc,Chang:1981qq,Djouadi:1987di})
The bare two loop mixed
QCD-EW contribution (with explicit inclusion of the tadpole) for the quark with mass $m_q$
reads:
\begin{eqnarray}
&& 
\Pi^{(2)}_{0,m_H^2,\alpha \alpha_s,q}  =  
\frac{g^2 g_s^2}{ (16 \pi^2)^2 }
N_c C_f \frac{m_q^2}{m_W^2}
\Biggl\{ 
- J_{0qq}(1,1,1;m_H^2) (n-3)
\nonumber \\ && 
+J_{0qq}(1,1,2;m_H^2) 
\left[
m_H^2 \!-\! 4  m_q^2 
\right] \frac{(n^2-5n+8)}{(n-4)(n-3)}
\nonumber \\ && 
+ A_0(m_q^2) B_0(m_q^2,m_q^2;m_H^2)
\frac{(n-2)}{(n-3)(n-4)}
\nonumber \\ && \hspace{5mm}
\times 
\left[ 
(n^3-8n^2+19n-16)
+ \frac{m_H^2}{2 m_q^2}
(n^2-5n+8)
\right]
\nonumber \\ && 
+ \left[ B_0(m_q^2,m_q^2;m_H^2) \right]^2
\left[ 
m_H^2 
\frac{(n-2)^2}{2(n-4)}
- m_q^2
\frac{2(n^2-4n+2)}{(n-4)}
\right]
\nonumber \\ && 
+ \left[ A_0(m_q^2) \right]^2 \frac{(n-2)(n^2-5n+8)}{2 m_q^2 (n-4)(n-3)}
+ \left[ A_0(m_q^2) \right]^2
\frac{3}{2} \frac{(n-1)(n-2)^2}{m_q^2 (n-3)}  
\Biggr\}
,
\end{eqnarray}
where the last terms come from the tadpole, $n$ is the dimension of space-time \cite{Hooft:1972fi} and  
\begin{eqnarray}
J_{0qq}(a,b,c;m^2)  & = &  
\left. 
\int \frac{d^n (k_1 k_2)}{[(k_1+k_2-p)^2]^a [k_1^2+m_q^2]^b [k_2^2+m_q^2]^c }  
\right|_{p^2=-m^2} ,
\nonumber \\
B_0(m_1^2,m_2^2,m^2) & = & 
\left. 
\int \frac{d^n k_1}{[k_1^2+m_1^2] [ (k_1-p)^2+m_2^2 ]} 
\right|_{p^2=-m^2} ,
\nonumber \\
A_0(m^2) & = & 
%\frac{\pi^{-n/2}}{\Gamma\left(3-\tfrac{n}{2}\right)}
\int \frac{d^n k_1}{k_1^2+m^2}  
\equiv \frac{4 (m^2)^{\frac{n}{2}-1}}{(n-2)(n-4)} . 
\end{eqnarray}
In accordance with Eq.~(\ref{mct}), 
the coefficient $\Delta_{m^2_H,\alpha_s \alpha,q}$  of order $O(\alpha \alpha_s)$ 
relating the pole and \MSb\ masses of the Higgs boson, $s_p-m_H^2$, 
can be written as 
\begin{eqnarray}
&& 
\Delta_{m_H^2, \alpha \alpha_s,q}
= 
\\ && 
\lim_{\ep \to 0 }
\Biggl( 
\frac{g^2 g_s^2}{(16 \pi^2)^2}
\left[ 
%\frac{1}{\ep} Z^{(2,1)}_{H,\alpha \alpha_s}
%\!+\! 
%\frac{1}{\ep^2} Z^{(2,2)}_{H,\alpha \alpha_s}
\frac{1}{\ep} Z^{(2,1)}_{H,\alpha \alpha_s,q}
\!+\! 
\frac{1}{\ep^2} Z^{(2,2)}_{H,\alpha \alpha_s,q}
\right]
\!-\! 
%Z_{m_t^2,\alpha_s} m_t^2 \frac{\partial }{\partial m_t^2} \Pi^{(1)}_{H,\alpha}
\frac{g_s^2}{16 \pi^2} \frac{1}{\ep}
Z_{m_q^2,\alpha_s} m_q^2 \frac{\partial }{\partial m_q^2} \Pi^{(1)}_{0,H,\alpha}
\!-\! 
%\Pi^{(2)}_{H, \alpha \alpha_s} 
\Pi^{(2)}_{0,m_H^2,\alpha \alpha_s,q}  
\Biggr) ,
\nonumber 
\label{DELTA}
\end{eqnarray}
where 
\begin{eqnarray}
\frac{\partial }{\partial m_q^2} \Pi^{(1)}_{0,H,\alpha}
& = &  
\frac{N_c}{m_W^2}
\frac{g^2}{16 \pi^2}  
\Biggl\{
B_0(m_q^2,m_q^2,m_H^2)
\left[ 
\frac{m_H^2\!-\!2m_q^2(n\!+\!1)}{2}
\right]
\!-\! 
\frac{(3n-2)}{2} A_0(m_q^2) 
\Biggr\}.
\nonumber \\
\end{eqnarray}
%As result of our calculation we find (for $m_b=0$): 
As result of our calculation we find: 
\begin{eqnarray}
Z_{H,\alpha \alpha_s,q}^{(2,1)} & = &  
%\frac{g^2 g_s^2}{(16 \pi^2)^2} N_c C_f  \frac{5}{4} \frac{m_t^2}{m_W^2} ,
\frac{g^2 g_s^2}{(16 \pi^2)^2} N_c C_f  \frac{5}{4} \frac{m_q^2}{m_W^2} ,
\quad 
Z_{H,\alpha \alpha_s,q} ^{(2,2)}  =   
- \frac{g^2 g_s^2}{(16 \pi^2)^2} N_c C_f  \frac{3}{2} \frac{m_q^2}{m_W^2} .
\end{eqnarray}
The contributions of other quarks with non-zero mass are additive. 
Exploring the $\ep$ expansion for the master integral $J_{0qq}$ from \cite{Davydychev:2003mv}, we have 
for t-quark contribution $(q=t)$:
\begin{eqnarray}
&& 
\Delta_{m^2_H,\alpha_s \alpha}     
\equiv 
\Delta_{m^2_H,\alpha_s \alpha,t}  =   
\frac{g_s^2 g^2}{ (16 \pi^2)^2} 
N_c C_f \frac{m_t^4}{m_W^2}
\Biggl\{ 
%\frac{4(1+y^2)(1+y)^2}{y(1-y)^2} 
\frac{4(z\!-\!2)(z\!-\!4)}{z} F(y)  
- 
\frac{4(1\!+\!y)^3}{y(1\!-\!y)} G(y)
\nonumber \\ && 
+ 
\frac{3+20y+16y^2-4y^3-9y^4}{2y(1-y)^2} \ln^2 y 
+ 
\frac{(1+y)}{(1-y)} \frac{(17+88y+17y^2)}{2y} \ln y 
\nonumber \\ && 
+ 
\frac{(131 + 258 y + 131 y^2)}{8y} 
- 6 \zeta_3 \frac{(1+y)^2(1+y^2)}{y(1-y)^2}
\Biggr\} 
\nonumber \\ &&
+ C_{H,\alpha \alpha_s}^{(2,2)} \ln^2 \left( \frac{m_t^2}{\mu^2} \right)
+ C_{H,\alpha \alpha_s}^{(2,1)} \ln   \left( \frac{m_t^2}{\mu^2} \right)
,
\label{DeltaH}
\end{eqnarray}
where
\begin{eqnarray}
z = \frac{m_H^2}{m_t^2} , \quad 
y = \frac{1-\sqrt{\frac{z}{z-4}}}{1+\sqrt{\frac{z}{z-4}}} , \quad 
z = - \frac{(1-y)^2}{y} , \quad 
4 m_t^2 - m_H^2 = m_t^2 \frac{(1+y)^2}{y} ,
\label{y}
\end{eqnarray}
and 
we have introduced the two functions $F(y)$ and $G(y)$ (see also \cite{Kniehl:1989yc,Djouadi:1994gf}) 
defined as\footnote{We cross checked, that Eq.~(\ref{DeltaH}) minus tadpole contribution
coincides with results of Ref.~\cite{Kniehl:1989yc,Kniehl:1993jc,Kniehl:1994ph,Djouadi:1993ss} after the following substitutions: 
\begin{eqnarray}
r = \frac{z}{4} , 
\quad 
1 - r = \frac{(1+y)^2}{4y} , 
\quad 
r_+ = 1/\sqrt{y} , 
\quad 
r_- = \sqrt{y} , 
\nonumber \\ 
f = - \frac{1}{2} \ln y , \quad 
g = \ln(1-y) - 1/2 \ln y , \quad
h = \ln(1+y) - 1/2 \ln y .
\end{eqnarray}
} 
\begin{eqnarray}
F(y) 
& = & 
3 \left[ \Li{3}{y} \!+\! 2 \Li{3}{-y} \right]
-
2 \ln y  \left[ \Li{2}{y} \!+\! 2 \Li{2}{-y} \right]
\nonumber \\ && \hspace{15mm}
- 
\frac{1}{2}
\ln^2 y \left[ \ln (1-y) \!+\! 2 \ln(1+y) \right] ,
\nonumber \\  
G(y) 
& = & 
\left[ \Li{2}{y} \!+\! 2 \Li{2}{-y} \right]
+
\ln y \left[ \ln (1-y) \!+\! 2 \ln(1+y) \right] ,
\end{eqnarray}
and 
\begin{equation}
\ln \left( \frac{m^2_H}{m_t^2} \right) = 2 \ln(1-y) - \ln y + i \pi .
\end{equation}
In Eq.~(\ref{DeltaH}) we explicitly factorized the RG logarithms, 
$C_{H,\alpha \alpha_s}^{(2,2)}$ 
and 
$
C_{H,\alpha \alpha_s}^{(2,1)}
$, which may be calculated also from the one-loop result and the mass anomalous dimensions 
(see \cite{Jegerlehner:2003wu,Jegerlehner:2004aj} for the general case).
From the parametrization
\begin{eqnarray}
M_H^2  & = & 
m_H^2 + \frac{g^2}{16 \pi^2} \left[ \Delta X_{H,\alpha}^{(1)} - C_{H,\alpha}^{(1)} L \right] 
+ \frac{g^2 g_s^2}{(16 \pi^2)^2}
\left[ 
\Delta X_{H,\alpha \alpha_s}^{(2)}
\!+\! 
C_{H,\alpha \alpha_s}^{(2,2)} L^2 
\!-\! C_{H,\alpha \alpha_s}^{(2,1)} L 
\right] 
\nonumber \\ & = & 
m_H^2 + \Delta_{m_H^2,\alpha}  + \Delta_{m_H^2,\alpha \alpha_s} 
\label{Delta-H}
, 
\end{eqnarray}
where $L = \ln \tfrac{\mu^2}{m_t^2}$, 
and using the fact that pole mass is RG invariant, 
we have:
\begin{eqnarray}
C_{H,\alpha}^{(1)} & = & m_H^2 Z_{H,\alpha}  , 
\quad 
\gamma_{m_t^2, \alpha_s} = Z_{m_t^2,\alpha_s}= - 6 C_f  ,
\\ 
2  C_{H,\alpha \alpha_s}^{(2,2)}
& = & 
Z_{m_t^2,\alpha_s} m_t^2 \frac{\partial }{\partial m_t^2} m_H^2 Z_{H,\alpha}
= - 3 m_H^2 C_f N_c \frac{m_t^2}{m_W^2}  ,
\\
C_{H,\alpha \alpha_s}^{(2,1)} 
& = &  
m_H^2 \gamma_{H,\alpha \alpha_s} 
+ Z_{m_t^2,\alpha_s} C_{H,\alpha}^{(1)}
+ Z_{m_t^2,\alpha_s} m_t^2 \frac{\partial }{\partial m_t^2} \Delta X_{H,\alpha}^{(1)} 
,
\end{eqnarray}
where 
\begin{equation}
Z_{H,\alpha}  =  
- \frac{3}{2} - \frac{3}{4} \frac{m_Z^2}{m_W^2} + \frac{3}{4} \frac{m_H^2}{m_W^2}
+ \sum_{\mbox{lepton}} \frac{1}{2} \frac{m_l^2}{m_W^2}
+  N_c \sum_{\mbox{u}} \frac{1}{2} \frac{m_u^2}{m_W^2}
+  N_c \sum_{\mbox{d}} \frac{1}{2} \frac{m_d^2}{m_W^2} .
\end{equation}
In terms of the variable $y$, defined by Eq.~(\ref{y}), the final result reads:
\begin{equation}
C_{H,\alpha \alpha_s}^{(2,1)}
= 
- 
C_f N_c 
\frac{m_t^4}{m_W^2}
\left[ 
   3 \frac{(1+y)(1+8y+y^2)}{y(1-y)} \ln y 
+  \frac{(17 + 38 y + 17 y^2)}{2y} 
\right] .
\end{equation}

%%%%%%%%%%%%%%%%%%%%%%%%%%%%%%%%%%%%%%%%%%%%%%%%%%%%%%%%%%%%%%%%%%%%%%%%%%%%%%%%%%%%%%%%%%%%%%%%%%%%%%
%%%%%%%%%%%%%%%%%%%%%%%%%%%%%%%%%%%%%%%%%%%%%%%%%%%%%%%%%%%%%%%%%%%%%%%%%%%%%%%%%%%%%%%%%%%%%%%%%%%%%%
%%%%%%%%%%%%%%%%%%%%%%%%%%%%%%%%%%%%%%%%%%%%%%%%%%%%%%%%%%%%%%%%%%%%%%%%%%%%%%%%%%%%%%%%%%%%%%%%%%%%%%
%%%%%%%%%%%%%%%%%%%%%%%%%%%%%%%%%%%%%%%%%%%%%%%%%%%%%%%%%%%%%%%%%%%%%%%%%%%%%%%%%%%%%%%%%%%%%%%%%%%%%%
%%%%%%%%%%%%%%%%%%%%%%%%%%%%%%%%%%%%%%%%%%%%%%%%%%%%%%%%%%%%%%%%%%%%%%%%%%%%%%%%%%%%%%%%%%%%%%%%%%%%%%
\subsection{$O(\alpha \alpha_s)$ corrections to the top Yukawa and Higgs self
  couplings}

The relation between the top Yukawa (Higgs) coupling and the Fermi constant $G_F$ is
obtained from Eqs.~(\ref{G}), (\ref{Delta-T}) and (\ref{Delta-H}) as:
\begin{eqnarray}
\frac{y_t^2(\mu^2)}{2 \sqrt{2}G_F M_t^2}
& = &  
\frac{m_t^2(\mu^2)}{M_t^2}  \frac{G_F(\mu^2)}{G_F} ,
\\
\frac{\lambda(\mu^2)}{\sqrt{2} G_F M_H^2}
%= \frac{\lambda(\mu^2)}{6} 
& = &  
\frac{m_H^2(\mu^2)}{M_H^2} \frac{G_F(\mu^2)}{G_F} ,
\end{eqnarray}
and the relation between the Higgs coupling constant $\lambda\equiv h_{\text{Sirlin}}$ used in \cite{Sirlin:1985ux} 
and the parametrization of \cite{Jegerlehner:2001fb,Jegerlehner:2002er,Jegerlehner:2002em,Jegerlehner:2003py} 
follows from the comparison of the RG functions: 
$
h_{\text{Sirlin}}  = \lambda_{\text{Jegerlehner}}(\mu^2)/6 
$.

The $O(\alpha \alpha_s)$ result for the top-Yukawa coupling reads (see Eq.~(21) in \cite{Jegerlehner:2003sp} and \cite{Kniehl:2004hfa})
\begin{eqnarray}
&& 
\sqrt{
\frac{y_t^2(\mu^2)}{ 2 \sqrt{2}G_F M_t^2} 
} - 1 
= 
\left( 
1 
+ \sigma_\alpha
+ \sigma_{\alpha_s}
+ \sigma_{\alpha \alpha_s}
\right)
\nonumber \\ && 
\times 
\left. 
\left(
1 
-  \Delta_{G_F,\alpha}
-  \Delta_{G_F,\alpha \alpha_s}
-  \sum_f \left[ m_f^2 \!-\! M_f^2 \right]_{\alpha_s} \frac{\partial}{\partial m_f^2} \Delta_{G_F,\alpha}
\right)^\frac{1}{2}  
\right|_{m_j^2 = M_J^2. e^2 = e^2_{OS}}
- 1 
\nonumber \\ && 
 =  
\left. 
\left( \sigma_\alpha \!-\! \frac{1}{2} \Delta_{G_F,\alpha} \!+\! \sigma_{\alpha_s } \right)
\right|_{m_j^2 = M_J^2. e^2 = e^2_{OS}}
\label{yukawa:12L}
\\ && 
+ 
\left. 
\left( 
\sigma_{\alpha \alpha_s} 
- \frac{1}{2} \Delta_{G_F,\alpha \alpha_s} 
- \frac{1}{2} \sigma_{\alpha_s } \Delta_{G_F,\alpha} 
- \frac{1}{2} \sum_f \left[ m_f^2 \!-\! M_f^2 \right]_{\alpha_s} \frac{\partial}{\partial m_f^2} \Delta_{G_F,\alpha}(m^2_t) 
\right)
\right|_{m_j^2 = M_J^2. e^2 = e^2_{OS}}
\nonumber
\end{eqnarray}
where $\sigma_X$ are defined in Eq.~(\ref{Delta-T}). 
The $O(\alpha \alpha_s)$ result for the Higgs coupling is 
\begin{eqnarray}
&& 
\frac{\lambda(\mu^2)}{\sqrt{2} G_F M_H^2} - 1 
=   
+ 
\left.
\left(
- \Delta_{G_F,\alpha} -  \frac{\Delta_{m^2_H,\alpha}}{M_H^2}
\right)
\right|_{m_j^2 = M_J^2. e^2 = e^2_{OS}}
\label{higgs:12L}
\\ &&
+ 
\left. 
\left( 
- \Delta_{G_F,\alpha \alpha_s} 
- \frac{\Delta_{m_H^2,\alpha \alpha_s}}{M_H^2}
- \left[ m_t^2 \!-\! M_t^2 \right]_{\alpha_s} \frac{\partial}{\partial m_t^2} 
\left[ 
\Delta_{G_F,\alpha} 
+ 
\frac{\Delta_{m_H^2,\alpha}}{M_H^2}
\right]
\right)
\right|_{m_j^2 = M_J^2. e^2 = e^2_{OS}}
\nonumber
,
\end{eqnarray}
where 
$$
\left[ m_t^2 \!-\! M_t^2 \right]_{\alpha_s}  = - 2 M_t^2 C_f \frac{g_s^2}{16 \pi^2} \left(4 - 3 \ln \frac{M_t^2}{\mu^2} \right) ,
$$
and the sum runs over all quarks. 

For completeness, we present also the explicit expressions for the derivatives
(for $N_c=3$, $C_F = 4/3$ and $m_b=0$):
\begin{eqnarray}
m_t^2 \frac{\partial}{\partial m_t^2}  \Delta_{m_H^2,\alpha} 
& =  & 
\frac{g^2}{16 \pi^2}
\frac{3 m_t^4}{m_H^2 m_W^2} 
\left[ 
\frac{1 \!+\! 4y \!+\! y^2}{y}
\!+\! \frac{(1 \!+\! y)(1 \!+\! 8y \!+\! y^2)}{2y(1 \!-\! y)}  \ln y 
\!+\! \frac{1}{2} \frac{m_H^2}{m_t^2} \ln \left( \frac{m_t^2}{\mu^2}  \right) 
\right]
,
\nonumber \\
\\ 
m_t^2 \frac{\partial}{\partial m_t^2} \Delta_{G_F,\alpha}  & =  & 
\frac{g^2}{16 \pi^2}
\Biggl\{
  \frac{m_t^4}{m_W^2 m_H^2} \left( 6 - 12 \ln \frac{m_t^2}{\mu^2} \right)
+ \frac{m_t^2}{m_W^2}    \left( \frac{3}{4} + \frac{3}{2} \ln \frac{m_t^2}{\mu^2} \right)
\Biggr\}
.
\end{eqnarray}
Terms of the order $O(\alpha)$, $O(\aS)$ in Eq.~(\ref{yukawa:12L}) and Eq.~(\ref{higgs:12L}) correspond to \cite{Hempfling:1994ar} and \cite{Sirlin:1985ux}, respectively.
Terms of the order $O(\alpha\aS)$ in Eq.~(\ref{yukawa:12L}) and Eq.~(\ref{higgs:12L}) are  the mixed electroweak-QCD
corrections
and 
$\Delta_{G_F,\alpha \alpha_s}$, 
$\sigma_{\alpha \alpha_s}$, 
$\Delta_{m_H^2,\alpha \alpha_s}$
are defined by Eq.~(\ref{G-alpha-alpha-s}), Eq.~(\ref{DeltaH}),
and Eq.(5.54) or Eq.(5.57) of \cite{Jegerlehner:2003py}.

For completeness we present also the the coefficient $\Delta_{m_H^2,\alpha}$. 
We divide all corrections into bosonic (diagrams without any fermions) 
and fermionic (diagrams exhibiting a fermion loop) ones: 
$\Delta_{m_H^2,\alpha} = 
\frac{g^2}{16\pi^2} m_H^2   
\left( 
\Delta_{m_H^2,\alpha,\mbox{boson}} 
+ 
\Delta_{m_H^2,\alpha,\mbox{fermion}}  
\right)
$. 
Using the notations of \cite{Jegerlehner:2001fb,Jegerlehner:2002er,Jegerlehner:2002em}
we may write the one-loop corrections in the following form
%\footnote{For simplicity we assume a diagonal Cabibbo-Kobayashi-Maskawa matrix.}
\begin{eqnarray}
\Delta_{m_H^2,\alpha,\mbox{boson}} 
& = & \frac{1}{2} - \frac{1}{2} \ln\frac{m_W^2}{\mu^2} - B(m_W^2,m_W^2;m_H^2)
\nonumber \\ 
& + & \frac{m_H^2}{m_W^2} 
\left( - \frac{3}{2} + \frac{9}{8} \frac{\pi}{\sqrt{3}} 
       + \frac{3}{8} \ln\frac{m_H^2}{\mu^2}
       + \frac{1}{4} B(m_W^2,m_W^2;m_H^2) + \frac{1}{8} B(m_Z^2,m_Z^2;m_H^2)  \right)
\nonumber \\ 
& + & \frac{m_Z^2}{m_W^2} \left( \frac{1}{4} - \frac{1}{4} \ln\frac{m_Z^2}{\mu^2}
       - \frac{1}{2} B(m_Z^2,m_Z;m_H^2) \right)
\nonumber \\ 
& + &  \frac{m_W^2}{m_H^2} 
\left( 3 - 3 \ln\frac{m_W^2}{\mu^2} + 3 B(m_W^2,m_W;m_H^2) \right)
\nonumber \\ 
& + & \frac{m_Z^4}{m_W^2 m_H^2} \left( \frac{3}{2} - \frac{3}{2} \ln\frac{M_Z^2}{\mu^2}
             + \frac{3}{2} B(m_Z^2,m_Z;m_H^2) \right) ,
\end{eqnarray}

\begin{eqnarray}
\Delta_{m_H^2,\alpha,\mbox{fermion}}  
& = &
\frac{1}{2} \frac{m_l^2}{m_W^2} \sum_{lepton} \Biggl[ B_0(m_l^2,m_l^2;m_H^2)
 \left( 1 - 4 \frac{m_l^2}{m_H^2} \right)
- 4 \frac{m_l^2}{m_H^2} \left(1 -  \ln \frac{m_l^2}{\mu^2} \right)
               \Biggr]
\nonumber \\
& + &  \frac{1}{2} \frac{m_q^2}{m_W^2} N_c\sum_{quark} \Biggl[
B_0(m_q^2,m_q^2;m_H^2 ) \left( 1 - 4 \frac{m_q^2}{m_H^2} \right)
-  4 \frac{m_q^2}{m_H^2} \left(1 -  \ln \frac{m_q^2}{\mu^2} \right)
                                                    \Biggr] ,
\nonumber \\
\end{eqnarray}

where (see Eq.~(E.6) in \cite{Davydychev:2003mv})
\begin{eqnarray}
B(m^2,m^2;m_H^2)
& = & \int\limits_0^1 dx \ln \Biggl( \frac{m^2}{\mu^2} x + \frac{m^2}{\mu^2} (1-x) - \frac{m_H^2}{\mu^2} x(1-x) - \mbox{i} 0 \Biggr) 
\nonumber \\
& = & 
\ln \frac{m^2}{\mu^2} - 2 - \frac{1+Y}{1-Y} \ln Y , 
\end{eqnarray}
with
$$
Y = \frac{1-\sqrt{\frac{r}{r-4}}}{1+\sqrt{\frac{r}{r-4}}} , 
\quad 
r = \frac{m_H^2}{m^2} .
$$
All results are collected in the Maple code of Ref.~\cite{MKL:pole}.

\begin{figure}
  \centering
  \includegraphics{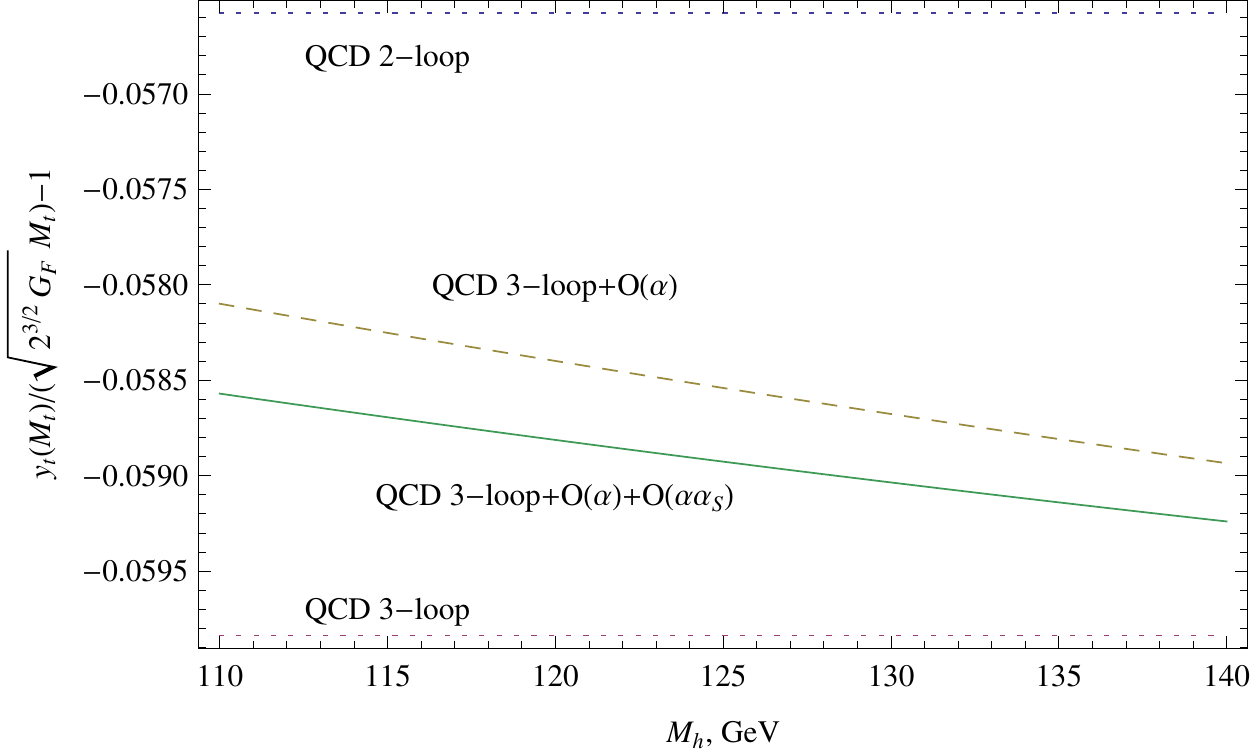}\\[2ex]
  \includegraphics{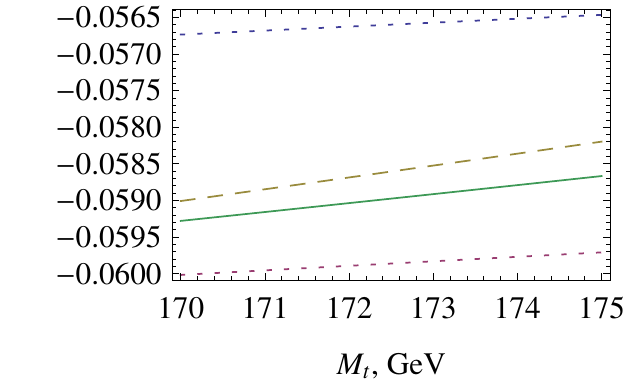} \includegraphics{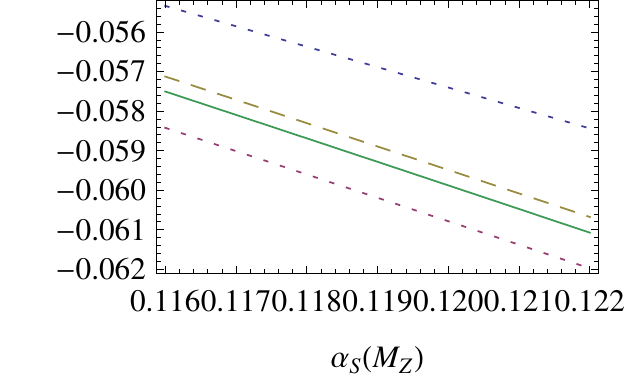}
  \caption{Contributions to the top Yukawa constant from QCD up to 2 loops, up to 3
    loops, QCD and 1 loop EW corrections $O(\alpha)$, and QCD with
    $O(\alpha)+O(\alpha\aS)$ corrections.  One parameter is vrying, the two others are
    chosen from $M_t=\unit[172.9]{GeV}$, $\aS=0.1184$, $M_h=\unit[125]{GeV}$.}
\end{figure}

\begin{figure}
  \centering
  \includegraphics{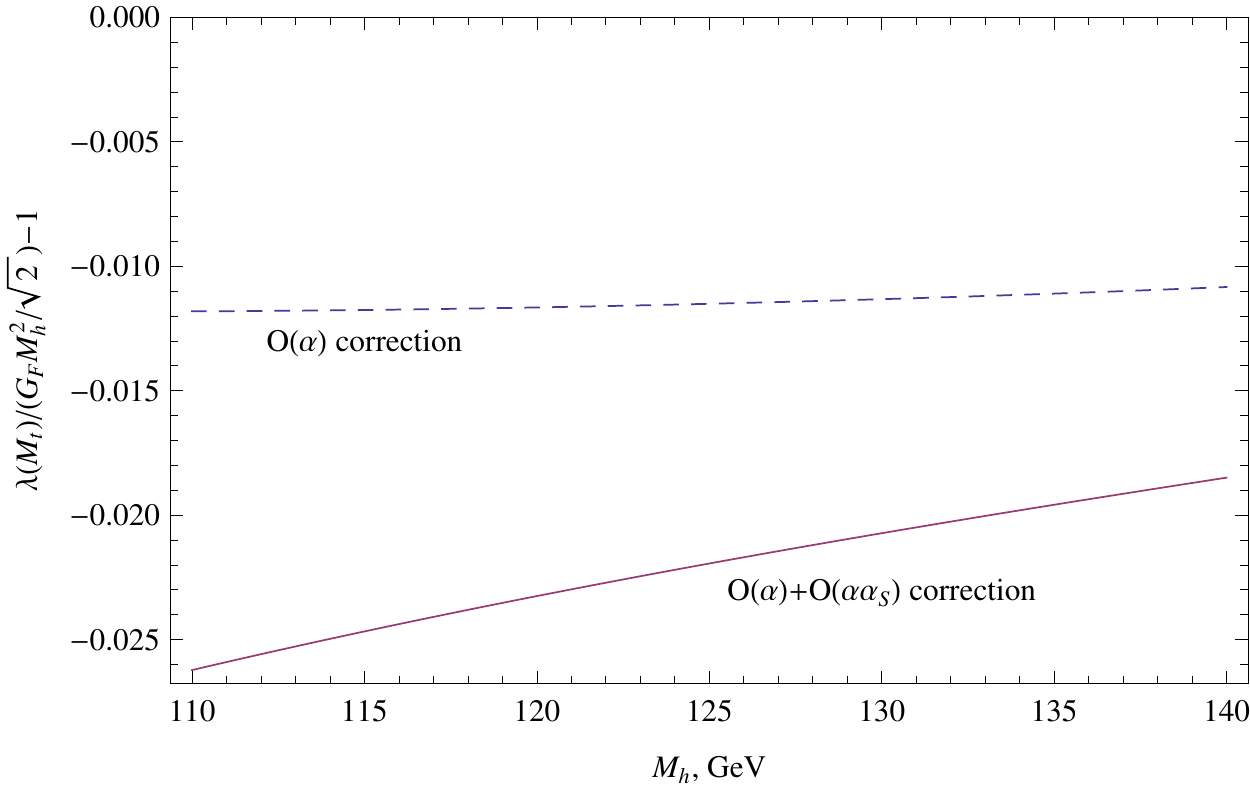}\\[2ex]
  \includegraphics{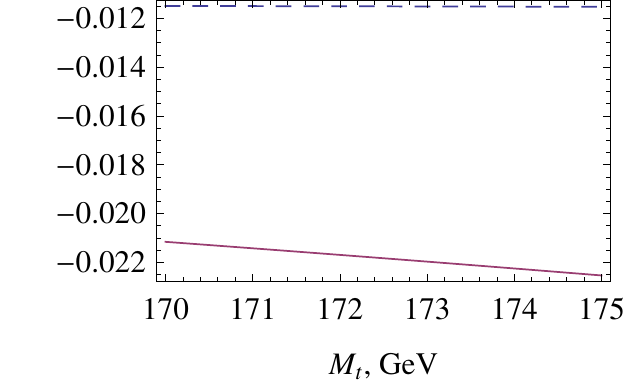} \includegraphics{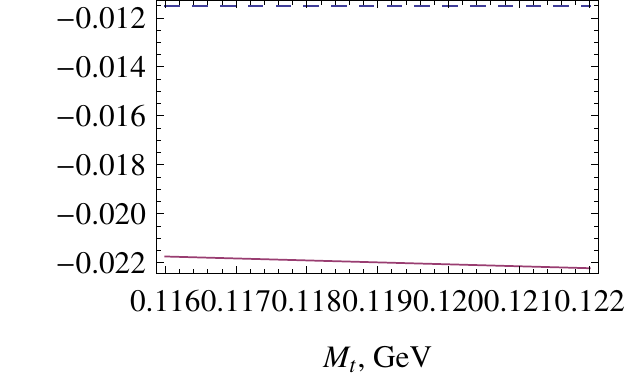}
  \caption{Contributions to the Higgs self coupling constant of the order $O(\alpha)$
    and $O(\alpha)+O(\alpha\aS)$.  One parameter is vrying, the two others are chosen
    from $M_t=\unit[172.9]{GeV}$, $\aS=0.1184$, $M_h=\unit[125]{GeV}$.}
\end{figure}

\begin{figure}
  \centering
  \includegraphics{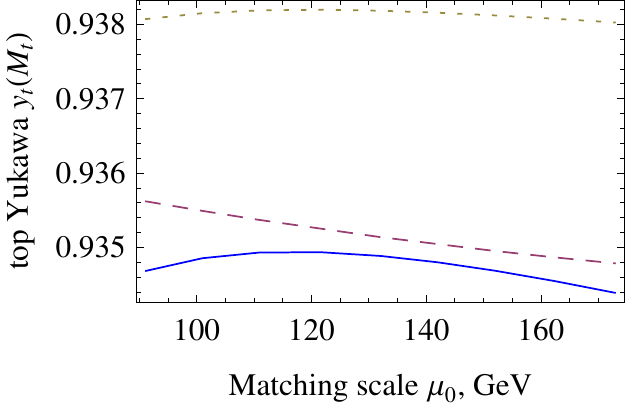}
  \includegraphics{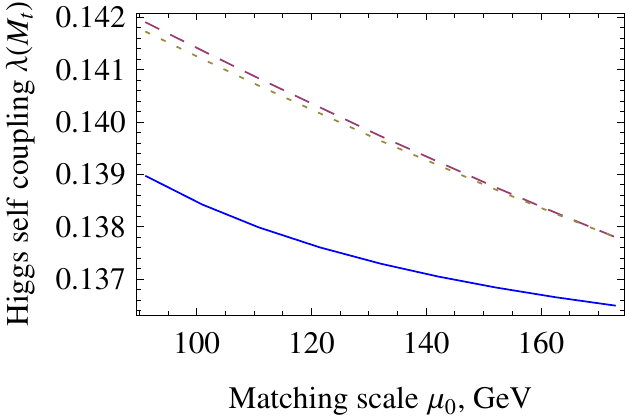}
  \caption{Top Yukawa (left) and Higgs coupling (right) at scale $M_t$.  The constants
    are extracted by using matching formulas at scale $\mu$ and then evolving to the
    scale $M_t$ by RG equations.  Blue solid line corresponds to using full
    $O(\alpha,\alpha\aS,\aS^3)$ matching formulas, dashed line is matching with
    $O(\alpha,\aS^3)$, dotted is matching with $O(\alpha,\aS^2)$.}
  \label{fig:MatchingTest}
\end{figure}

\section{Beta functions}

Two loop SM beta functions above the top mass are collected in~\cite{Espinosa:2007qp}
(see~\cite{Fischler:1982du,Fischler:1983sf,Ford:1992mv,Jack:1982sn,Jack:1982sr,
  Jack:1984bg,Jack:1984vj,Machacek:1983tz,Machacek1984,Machacek1985} for original
works).  The three loop beta functions can be read off
\cite{Mihaila:2012fm,Chetyrkin:2012rz}.

Below the top mass the one loop beta functions for the gauge couplings were used to
evolve the PDG values from $M_Z$ to $M_t$.  For example, for the $\alpha(\mu)$
\begin{align}
  \amu    &= \frac{\amz}{1+\frac{11}{6\pi}\amz\log(\frac{\mu}{m_Z})}.
\end{align}
The higher loop corrections are not important numerically for the electroweak
constants for the small energy range between $M_Z$ and $M_t$.

For the strong coupling $\aS\equiv g_S^2/(4\pi)$ the RG equation up to order
$O(\aS^3)$ is used
\begin{align}
  \label{eq:baSQCD}
  \frac{d\aS}{d\log\mu} = - (11-\frac{2}{3}N_f) \frac{\aS^2}{2\pi}
  - (51-\frac{19}{3}N_f) \frac{\aS^3}{4\pi^2}
  - (2857-\frac{5033}{9}N_f+\frac{325}{27}N_f^2) \frac{\aS^4}{64\pi^3},
\end{align}
and $N_f=5$ is the number of flavors below the top quark.  Strictly speaking, the value
of $\aS(M_t)$ obtained from this equation should be also shifted to the 6-quark
value by
\begin{equation}
  \label{eq:2}
  \alpha_{S,N_f=6}(M_t) =
  \alpha_{S,N_f=5}(M_t)-\frac{11}{72\pi^2}\alpha_{S,N_f=5}^3(M_t),
\end{equation}
but this introduces a negligible effect ($<\unit[0.1]{GeV}$) for the Higgs mass.

In all the formulas of the Appendix \ref{app:PoleMatching} we use the values of
$\alpha$ and $\aS$ at the matching scale $\mu$.

%%% Local Variables: 
%%% mode: latex
%%% TeX-master: "higgs126"
%%% End: 

%\bibliographystyle{JHEP}
%\bibliography{all,local}
\providecommand{\href}[2]{#2}\begingroup\raggedright\endgroup

\end{document}